\documentclass[12pt]{article} 

\usepackage{amssymb,amsmath, amsfonts}
\usepackage{cases}
\usepackage{latexsym}
\usepackage{graphicx}
\usepackage{relsize}
\usepackage{natbib}
\usepackage{rotating}
\usepackage[includeheadfoot,left=1.in,right=1.in,top=1.in,bottom=1.in,bindingoffset=0.in,nohead]{geometry}
\usepackage[onehalfspacing]{setspace}
\usepackage{booktabs}
\usepackage{appendix}
\usepackage[pdfborder={0 0 0}]{hyperref}
\usepackage{float}
\usepackage{xcolor}
\usepackage{caption}
\usepackage{subcaption}
\usepackage{minibox}

\usepackage{floatpag}

\newcommand{\argmin}{\arg\,\min}

\begin{document}

\title{On the Aggregation of Probability Assessments: \\Regularized Mixtures of Predictive Densities for \\Eurozone Inflation and Real Interest Rates}

\author{Francis X. Diebold\\University of Pennsylvania \and Minchul Shin\\
Federal Reserve Bank of  Philadelphia \and Boyuan Zhang \\ University of Pennsylvania\\$~$ \vspace{-0.2in}}

\maketitle

\thispagestyle{empty}

\begin{spacing}{1}

\vspace{-0.1in}

 \noindent \textbf{Abstract}: We propose methods for constructing regularized mixtures of density forecasts.  We explore a variety of objectives and regularization penalties, and we use them in a substantive exploration of Eurozone inflation and real interest rate density forecasts. All individual inflation forecasters (even the ex post best  forecaster) are outperformed by our regularized mixtures. 
 From the Great Recession onward, the optimal regularization tends to move density forecasts' probability mass from the centers to the tails, correcting for overconfidence.

%
%
%
%
%
%

\bigskip

%
%
%
%
%

\noindent {\bf Acknowledgments}: For helpful comments and/or assistance we are grateful to Umut Akovali, Brendan Beare, Graham Elliott, Rob Engle, Domenico Giannone, Christian Hansen, Nour Meddahi, Mike McCracken, Marcelo Medeiros, James Mitchell, Joon Park,  Hashem Pesaran, Youngki Shin, Mike West, and Ken Wolpin. We are also grateful to conference participants at $\text{EC}^2$, and seminar participants at KAEA and AMLEDS.  The views expressed in this paper are solely those of the authors and do not necessarily reflect the views of the Federal Reserve Bank of Philadelphia or the Federal Reserve System.
 

\bigskip


\bigskip


\bigskip

{\noindent {\bf Contact}:  fdiebold@sas.upenn.edu, minchul.shin@phil.frb.org}

\end{spacing}

%
%
%
%

\newpage

\setcounter{page}{1}
\thispagestyle{empty}

\section{Introduction} \label{intro}

Forecast combination for a  series $y$ involves transforming a set of forecasts of $y$,  $f = (f_{1}, ..., f_{K})'$,  into a ``combined", and hopefully superior, forecast $c(f)$.  Most of the huge literature focuses on linear combinations of univariate point forecasts, in which case we can write the combined forecast as 
$c(f; \omega) = \omega ' f$,
 for combining weight vector $\omega = (\omega_1, ...,  \omega_K)'$.\footnote{Broad and insightful surveys include \cite{timmermann2006}, \cite{ET2016}, and \cite{RePEc:tin:wpaper:20180069}.}  We typically proceed under quadratic loss, choosing the weights to minimize the sum of squared combined forecast errors ($SSE$),   
$$
SSE(c(f; \omega), y) = \sum_{t=1}^T \left (y_t - \omega ' f_t \right )^2 ,
$$
where the sample of forecasts and realizations covers $t=1, ..., T$. That is, we simply run the least-squares regression  $y \rightarrow f_{1}, ..., f_{K}$, so that\footnote{We assume unbiased forecasts, so there is no need for an intercept.}
$$
\hat{\omega} = \arg \min_{\omega} \Big (SSE(c(f; \omega), y)   ~\Big ).
$$
This is the classic \cite{BatesandGranger1969} and \cite{granger1984} solution.

Recent point forecast combination literature such as \cite{diebold2018}, however, focuses instead on weights that solve a \textit{penalized} estimation problem,
\begin{equation} \label{basic}
\hat{\omega} = \arg \min_{\omega} \Big ( ~ Objective(c(f; \omega), y)  ~+~ \lambda \cdot  Penalty(\omega)  ~\Big ),
\end{equation}
where the Lagrange multiplier $\lambda$ governs the strength of the penalty. Maintaining quadratic loss we have
$$
\hat{\omega} = \arg \min_{\omega} \Big ( ~ SSE(c(f; \omega), y)   ~+~ \lambda \cdot  Penalty(\omega)  ~\Big ).
$$
If $\lambda \mathord =0$ we obviously obtain the Bates-Granger-Ramanathan solution, but the recent literature focuses on  $\lambda \mathord > 0$. This produces regularization, which can be highly valuable in the finite samples often of practical relevance, particularly for economic survey forecasts where the sample size $T$ is often very small relative to the number of forecasters $K$. The precise form of the penalty determines the precise form of regularization, but in general it involves selection and/or shrinkage in directions guided by the penalty.  For example, the famous LASSO penalty of  \cite{tibshirani1996}, $Penalty(\omega) =  \sum_{k=1}^K  |{\omega_k} |$, induces both selection to 0 and shrinkage toward 0.

In this paper we extend the idea of regularized forecast combination to the density forecast case.  Density forecasting is important because predictive densities are complete probabilistic statements, which are always desirable, sometimes invaluable, and increasingly available.  Density forecasts provide much more information, for example, than interval forecasts, which in turn provide more information than point forecasts.\footnote{The evaluation of interval forecasts, moreover, is fundamentally problematic, as detailed in recent work by  \cite{adss2018} and \cite{Gneiting2020}.}  

 We work with  ``linear opinion pools" (mixtures), as in the key contributions of \cite{Hall2007}, \cite{GewekeAmisano2011} and \cite{amisano2017}, but we consider a variety of estimation objectives,  and most importantly, we introduce regularization constraints.   Our regularized density forecast combinations are regularized mixtures, and  important subtleties arise in constructing appropriate penalties for mixture regularization.  In this paper we confront this situation and propose several solutions.
 
 Our methods are related to earlier and current work in both the econometrics and statistics literatures.   A basic insight underlying our work and much of the recent literature is that Bayesian model averaging (BMA) as traditionally implemented is unattractive for combining density forecasts from misspecified models, because it fails to acknowledge misspecification \citep{Diebold1991}.  That is, it assumes implicitly or explicitly   that one of the models is ``true", in which case the posterior predictive density asymptotically puts all probability on that model, so that BMA actually \textit{fails} to average. 
  Instead, once we acknowledge that all models are misspecified,  we  want a method capable of delivering a defensible and \textit{diversified} portfolio (weighted average) of models, even asymptotically.
 
  In one strand of  econometrics literature this led \cite{Hall2007},  \cite{brodie2009},  \cite{GewekeAmisano2011}, and \cite{amisano2017} \textit{inter alia} to move away from BMA, working instead with linear opinion pools that optimize the log score.  In a different  strand of econometrics literature  that also moved  away from BMA, it led   \cite{billio2013} to treat density forecast combination  as a nonlinear filtering problem, potentially with time-varying mixture weights.   Parallel developments in the statistics literature now acknowledge misspecification, distinguishing between ``M-open" vs. ``M-complete" situations, and achieve diversified density forecast mixtures  by ``stacking" predictive densities \citep{yao2018}, or via ``dynamic Bayesian predictive synthesis" \citep{McAlinnWest2017bpsJOE}.

We pick up from there and proceed as follows. In section \ref{OandP1} we discuss  objectives  for mixture regularization, that is, various choices and issues associated with  $Objective(c(f; \omega), y)$.  Then in section \ref{OandP2} we treat choices and issues associated with 
$Penalty(\omega)$, starting with the key unit simplex penalty, which we maintain throughout, and then introducing  hybrid penalties that blend the simplex penalty with others.  In section \ref{MonteCarlo} we present Monte Carlo evidence on the efficacy of our procedures.  In section  \ref{empirical} we present empirical results for European Central Bank (ECB) survey density forecasts of Eurozone inflation and real interest rates.  We conclude in section \ref{concl}.

\section{Objectives} \label{OandP1}

   Consider a discrete density (histogram) forecast  for a scalar variable $y$, which takes values in $m = 1, ..., M$ bins, or categories.\footnote{We focus largely on the discrete case, because it is the one of practical relevance for survey forecasts that we eventually analyze.  Parallel developments of course exist for the continuous case.} Denote the forecast by $p = (p_1, ..., p_M)'$. We start with density forecast ``scores" for a single forecaster in a single period in sections \ref{score1}-\ref{score3}, we extend the discussion to multiple forecasters and periods in section \ref{score5}, and we provide additional  discussion in section \ref{score6}.

  \subsection{Log Score } \label{score1}
  
  The log score \citep{Good1952, WinklerMurphy1968} is
  \begin{equation}  \label{logsc}
    L(p, y) =  -  \log \left ( \sum_{m=1}^M  {p}_{m}\, 1(y \in b_m)  \right ),
  \end{equation}
  where  $p_{m}$ is the probability assigned to bin $b_m$,  and $1(y \in b_m) = 1$ if $y \in  b_m$ and 0 otherwise.

  Ranking density forecasts by  $L$, where smaller is  better, reflects a preference for ``small surprises". In a frequentist interpretation, $L$ is just the  (negative of the) log predictive density evaluated at the realization; that is, it is the (negative of the) predictive log likelihood. In a Bayesian interpretation,  $L$ is, desirably,  a strictly proper scoring rule.\footnote{On scoring rules see \cite{gneiting2007} and the references therein.}

       \subsection{Brier Score} \label{score2}

 The  Brier score \citep{Brier1950} is:  
      
      $$
      B(p, y)  = \frac{1}{M} \sum_{m=1}^M  \left ( p_{m}  - 1(y \in b_m) \right ) ^2.
      $$
 The Brier score generalizes the idea of quadratic loss to density forecasts. Indeed $B$ is effectively the same as the so-called  ``quadratic score",    
      \begin{equation} \label{Qeqn}
       Q(p, y)  = -2  \left ( \sum_{m=1}^M  {p}_{m} \, 1(y \in b_m)  \right ) + \left ( \sum_{m=1}^M  {p}_{m}^2   \right ),
      \end{equation}
        as noted by \cite{czado2009}.  Rankings by $Q$ must match rankings by $B$, because one is a positive monotonic transformation of the other. Both $B$ and $Q$ are strictly proper scoring rules under weak conditions.

        \subsection{Ranked  Score} \label{score3}
      
        The ranked score \citep{Epstein1969} is,         
      $$
                R(p, y)  =    \sum_{m=1}^{M}   \left ( P_{m} - 1(y \le b_{m+}) \right )^{2},
      $$
                where $P_{m} = \sum_{h=1}^{m} p_{}(b_{h})$ is the cdf of the   
                density forecast $p$, defined on bins $b_m  = [b_{m-}, b_{m+}]$, $m=1, ..., M$.  $R$ effectively proceeds by comparing realizations to the cdf forecast rather than the density forecast. $R$ is strictly proper under weak conditions.

\subsection{Multiple Forecasters and Time Periods}  \label{score5}
 
Let us now modify the notation to identify the specific forecaster, $k$. Thus far there has been no need, as we have considered just one forecaster, but shortly we will want to consider a set of forecasters, $k=1, ..., K$. This is just a notational change, inserting ``$k$" subscripts in the relevant places. In addition let us write the scores for a set of periods, $t=1, ..., T$, rather than for just one period. This just involves summing over time.

We have:
$$
L_k(\bold{p_k}, \bold{y}) =  \sum_{t=1}^T \left (- \log \left ( \sum_{m=1}^M  {p}_{mkt}\, 1(y_t \in b_m)  \right )   \right ), ~ k=1, ..., K
$$

$$
B_k(\bold{p_k}, \bold{y}) =  \sum_{t=1}^T \left ( \frac{1}{M}  \sum_{m=1}^M  \big ( p_{mkt}  - 1(y_t \in b_m) \big ) ^2  \right ), ~ k=1, ..., K
$$

$$
R_k(\bold{p_k}, \bold{y}) =  \sum_{t=1}^T  \left (    \sum_{m=1}^{M} \Big ( P_{mkt} - 1(y_t \le b_{m+}) \Big )^{2}  \right ), ~ k=1, ..., K,
$$
where $\bold{p_k} = (p_{k1}, ..., p_{kT})$ is the sequence of density forecasts over time for forecaster $k$, and  $\bold{y} = (y_1, ..., y_T)$ is the sequence of realizations over time.

\subsection{Discussion}  \label{score6}

Thus far we have implicitly emphasized the differences among the $L$, $B$, and $R$ scores, but there are also many similarities. 

$B$, for example, might appear linked to Gaussian environments, because it is a mean-squared error analog, unlike $L$ which is based directly on the likelihood and therefore  valid under great generality.  But it is not; indeed its ``$Q$ version" (\ref{Qeqn}),
$$
Q = -2   L  + \left ( \sum_{m=1}^M  {p}_{m}^2   \right ),
$$
reveals its close link to  $L$. Moreover, $B$ remains a strictly proper scoring rule regardless of distributional environment.

Now consider $R$. First, it is interesting to note that $R$ is a generalization of absolute-error loss to density forecasts, just as $B$ is a generalization of squared-error loss to density forecasts.  In particular, \cite{gneiting2007} show that $R$ is driven by  $E_p |Y-y|$:
$$
R(p, y) = E_p |Y-y| - \frac{1}{2} E_p |Y - Y'|,
$$
where $Y$ and $Y'$ are independent copies of a random variable with distribution $p$.

Second, $R$'s generalization of absolute-error loss ($MAE$) to density forecasts also makes it a generalization of the  \cite{DieboldShin2017} stochastic error distance ($SED$), because $MAE$ and  $SED$ rankings must agree, and interestingly,  $SED$ is based on cdf divergences, just as is $R$.

Finally, although $R$ might appear linked to a particular  (Laplace) distributional environment, because it is an absolute-error analog, it is not.  $R$ is a strictly proper scoring rule regardless of  distributional environment.

\section{Penalties} \label{OandP2}

Our goal is to produce  mixtures of density forecasts,                 
$$
c(\omega) = \sum_{k=1}^K  \omega_k p_{k},
$$
with regularized mixture weights $\omega = (\omega_1, ..., \omega_K)'$.          We score mixtures in the same way as we scored individual density forecasts.  The only difference is that we now score the mixture, $c(\omega) $, rather than an individual forecast, $p_{k}$. 

Thus far we have focused on  appropriate objectives for regularized mixture weight estimation,  $objective( c(\omega) , y)$, and we emphasized use of strictly proper density forecast scoring rules. 
Now we consider appropriate constraints for regularized mixture weight estimation, $penalty( \omega)$.  As we shall see, imposition of the unit simplex constraint (i.e., imposing that mixture weights be non-negative and sum to one: $\omega_i {\ge} 0~ \forall i$ and $\sum_{i=1}^K \omega_i =1$) provides essential regularization.  In addition, however, simultaneous imposition of other regularization constraints may also be helpful.

\subsection{Simplex}

The unit simplex constraint has two parts: non-negativity and sum-to-one.  For point forecasts we can relax both parts and potentially achieve better combined point-forecasting performance, as recognized by \cite{granger1984} and done routinely ever since.  As first recognized in the pioneering work of  \cite{brodie2009}, it turns out that density forecasts are different: \textit{When combining density forecasts it is crucial to impose (both parts of) the simplex constraint}.  

First consider non-negativity. For point forecasts, allowing negative combining weights can improve performance, in a fashion analogous to allowing short positions in a financial asset portfolio. For density forecasts, in contrast, negative weights are unambiguously problematic, producing pathologies even if sum-to-one holds, because negative mixture weights can drive parts of the mixture density negative.

Now consider sum-to-one. Immediately, sum-to-one is required for the mixture combination to be  a valid probability density.\footnote{See  also  \cite{yao2018}, who  briefly discuss issues related to the imposition of convex mixture weights.}  Moreover, and separately, the solution to the mixture weight estimation problem can be pathological without imposition of sum-to-one. To see this, consider a simple example with two continuous density forecasts and a log score objective. We have
\[
\hat{\omega} = \arg \min_{\omega_{1}, \omega_{2}}  \left ( - \sum_{t=1} ^T \log (\omega_{1} f_{1,t}(y_{t}) + \omega_{2} f_{2,t}(y_{t})) \right ),
\]
where $f_{k,t}(y_{t})$ is forecaster $i$'s density forecast evaluated at the realization, $y_{t}$. Without the sum-to-one constraint,  the optimal solution is not well defined:  either $\omega_{1} {\rightarrow} \infty$ or $\omega_{2} {\rightarrow} \infty$ leads to the smallest possible objective function value, because $f_{1,t}$ and $f_{2,t}$ are non-negative for any $y_{t}$.

For all of the above reasons, we henceforth impose both the non-negativity and sum-to-one parts of the simplex constraint.  Interestingly, moreover, their imposition is not only necessary to eliminate pathologies, but also desirable to provide regularization.  In particular, the simplex constraint clearly imposes a particular $L^1$ ``parameter budget"; it is effectively a special case of LASSO.

Assembling everything, the  basic regularized estimator with log score objective \citep{GewekeAmisano2011,amisano2017} is\footnote{Other objectives may of course be used, as discussed earlier  in section \ref{OandP1}. Note that for a histogram forecast we have $f_{k,t}(y_{t}) = \sum_{m=1}^{M} p_{mkt} 1 (y_{t} \in b_{m})$. } 
\begin{equation}  \label{AG}
\arg \min_{\omega} \left ( {-\sum_{t=1}^{T} \log \left( \sum_{k=1}^{K} \omega_{k} f_{k,t}(y_{t})\right)}_{\text{ }} \right )
\end{equation}
\begin{equation*}
\text{s.t.}~ \omega_{k} \in (0,1), ~ \sum_{k=1}^{K}\omega_{k} = 1.
\end{equation*}
The methodological question remains, however, of how to provide additional, and more flexible, regularization, as does the substantive situation-specific empirical question of whether and where additional regularization is helpful.  In the remainder of this paper we work toward answering both questions.

\subsection{Simplex+Ridge}  \label{meth1}

$L^1$ simplex regularization is a special case of $L^1$ LASSO regularization, corresponding to a specific choice of LASSO regularization parameter.  Hence we cannot introduce additional $L^1$ regularization.  

Additional regularization of some other type may nevertheless be useful for a variety of reasons.  One reason is that the sparsity promoted by the simplex constraint may not be desirable \citep{Illusion}, so we may want to shrink all $K$ mixture weights away from 0, thereby  ``undoing" the selection implicit in the LASSO-style $L^1$ penalty, allowing for non-zero mixture weights on all forecasts.  We focus in particular on introducing shrinkage toward an equally-weighted mixture (i.e., shrinkage of  all $K$  weights toward $1/K$).

Consider, for example, introducing $L^2$ regularization.  Immediately, incorporating an $L^{2}$ penalty in addition to the simplex constraint, we have:\footnote{For transparency we make most of our arguments using a log score objective.}
\begin{equation}
\label{cup} 
\hat{\omega} = \arg \min_{\omega} \left ( \underbrace{-\sum_{t=1}^{T} \log \left( \sum_{k=1}^{K} \omega_{k} f_{k,t}(y_{t})\right)}_{\text{log score}}~+ ~\underbrace{\lambda \left(\sum_{k=1}^{K} \left(\omega_{k}-\frac{1}{K}\right)^{2}\right)}_{L^2~\text{penalty}} \right ) 
\end{equation}
\begin{equation*}
\text{s.t. } \omega_{k} \in [0,1], ~ \sum_{k=1}^{K}\omega_{k} = 1.
\end{equation*}
This parallels the egalitarian ridge estimator of \cite{diebold2018}, with an additional simplex constraint imposed.  Note that, due to the simplex constraint, the solution may discard some forecasters (setting some weights approximately  if not exactly to zero), but that situation becomes progressively less likely as $\lambda$ grows, pulling the weights toward equality.

 We can re-write (\ref{cup}) as
\begin{equation} \label{glas2} \small
\hat{\omega} =  \arg \min_{\omega} 
\left (
 \underbrace{-\sum_{t=1}^{T} \log \left( \sum_{k=1}^{K} \omega_{k} f_{k,t}(y_{t})\right)}_{\text{log score}}
  ~+ \underbrace{\lambda_{1} \left(\sum_{k=1}^{K} | \omega_{k} | -1\right)}_{L^1~\text{simplex/LASSO~penalty}}   
          +~~ \underbrace{\lambda_{2} \left(\sum_{k=1}^{K} \left(\omega_{k}-\frac{1}{K}\right)^{2}\right)}_{L^2~\text{ridge ~penalty}} 
  \right ),  
\end{equation}
\begin{equation*} 
 \text{s.t. } \omega_{k} \in [0,1],
\end{equation*}
which emphasizes that simplex+ridge regularization involves  a combination of  $L^{1}$ and $L^{2}$ penalties.\footnote{Equation (\ref{glas2}) also reveals that simplex+ridge is closely related  to an additive-penalty version of partial egalitarian LASSO  \citep{diebold2018}, but with the egalitarian penalty done in $L^2$ (ridge) form rather than $L^1$ (LASSO) form.} Note, however,  that we are not free to choose $\lambda_{1}$, because the sum-to-one constraint must bind; equations (\ref{cup}) and (\ref{glas2}) instead coincide for ``large enough" $\lambda_1$.

Equation (\ref{glas2}) in turn reveals that simplex+ridge regularization is closely related to the elastic net of \cite{zh2005}.  The elastic net penalty is  
$$
Penalty(\omega) ~~= \underbrace{\alpha \sum_{k=1}^K     |{\omega_k}|}_{L^1~\text{LASSO ~penalty}}    + ~~~\underbrace{(1 {-} \alpha) \sum_{k=1}^K   \omega_k^2}_{L^2~\text{ridge ~penalty}},
$$ 
where $\alpha {\in} [0,1]$ is a parameter, so that elastic net also involves  combinations of  $L^{1}$ and $L^{2}$ (that is, LASSO/simplex and ridge) penalties. Elastic net is well known to work well for regularization problems with many correlated predictors, exactly the situation of relevance for the large sets of economic forecasts on which we focus. 

\subsection{Simplex+Divergence}

Here we move from simplex+ridge to simplex plus a general penalty based on the divergence between two discrete probability measures.  As we will see, the divergence penalty includes simplex+ridge as a special case, but it  also introduces a rich variety of new possibilities.  Write the estimator as
\begin{equation}
\label{gennn} 
\hat{\omega} = \arg \min_{\omega} \left ( \underbrace{-\sum_{t=1}^{T} \log \left( \sum_{k=1}^{K} \omega_{k} f_{k,t}(y_{t})\right)}_{\text{log score}} + \underbrace{\lambda \, D\left(\omega,\omega^{*}\right)}_{\text{penalty}} \right ) 
\end{equation}
\begin{equation*}
\text{s.t. } \omega_{k} \in [0,1],~  \sum_{k=1}^{K}\omega_{k} = 1,
\end{equation*}
where $D(\omega,\omega^{*})$ is a measure of divergence between $w$ and $w^{*}$. The key insight is that once the simplex restriction is imposed, $\omega$ can be interpreted as a discrete probability measure on $\{1,2,...,K\}$.  If we let $\omega^{*}$ be the uniform probability mass function with weight $1/K$ on each outcome, then the penalized optimization (\ref{gennn}) shrinks the solution toward equal weights.

Maintaining  uniform $\omega^{*}$ throughout,  but using different divergence measures $D(\omega,\omega^{*})$, we obtain  new regularized estimators. For example:

\begin{enumerate}
        
        \item The $L^{2}$ norm, 
        $$
        D(\omega, \omega^{*}) = \sum_{k=1}^{K} \left(\omega_{k}-\frac{1}{K}\right)^{2},
        $$
        produces the simplex plus egalitarian ridge penalty given in (\ref{cup}) and (\ref{glas2}).
        
\item The $L^1$ norm (total variation), 
$$
D(\omega, \omega^{*}) = \sum_{k=1}^{K} \left  |\omega_{k} - \frac{1}{K} \right |,
$$
 produces 
 a simplex plus egalitarian LASSO  penalty  \citep{diebold2018}.

\item  Kullback-Leibler divergence (entropy) from $\omega$ to $\omega^*$, 
$$
D(\omega, \omega^{*})= - \log K  - \sum_{k=1}^{K} \log \omega_k,
$$ produces a ``simplex+entropy" penalty, $- \sum_{k=1}^{K} \log \omega_k $.  In Appendix \ref{derivation} we  formally show that the  simplex+entropy regularized estimator,
\begin{equation}
\label{log2}
\hat{\omega} = \arg \min_{\omega} \left ( \underbrace{-\sum_{t=1}^{T} \log \left( \sum_{k=1}^{K} \omega_{k} f_{k,t}(y_{t})\right)}_{\text{log score}}+ \underbrace{\lambda \left(-\sum_{k=1}^{K} \log(\omega_{k})\right)}_{\text{entropy penalty} } \right )
\end{equation}
\begin{equation*}
\text{s.t.}~ \omega_{k} \in (0,1), ~ \sum_{k=1}^{K}\omega_{k} = 1,
\end{equation*}
 arises as the posterior mode in a Bayesian analysis with  a log score (pseudo-)likelihood and a Dirichlet prior, which puts positive probability only on the unit simplex and also shrinks weights toward equality for a certain hyperparameter configuration.

\item  R\'{e}nyi divergence of order $\alpha$ from $\omega$ to $\omega^{*}$, 
\[
D_{\alpha}(\omega^{*}||\omega) = \frac{1}{\alpha-1} \log \left(\sum_{k=1}^{K} \frac{1/K^{\alpha}}{\omega_{k}^{\alpha-1}}\right),
\]
encompasses various statistical divergences including Kullback-Leibler divergence ($\alpha =1$) and Hellinger distance ($\alpha=2$), and can be used to produce still more interesting regularized estimators.\footnote{ R\'{e}nyi divergence, moreover,   is  equivalent to  Cressie-Read discrepancy up to an affine transformation.} 
\end{enumerate}
\noindent All of the above  divergence functions shrink the density mixture weights toward equality, thereby promoting inclusion of  more forecasters in the regularized mixture.  Importantly, the optimization that defines the  regularized estimator  (\ref{gennn}) is convex so long as $D(\omega, \omega^{*})$ is a convex function of $\omega$, because  the log score and simplex constraints are convex functions of $\omega$.  This  makes numerical computation of the estimator straightforward.

\subsection{Partially Egalitarian LASSO and Subset Averaging} \label{meth2}

One might want a density forecast version of partially egalitarian penalization, as developed for the point forecast case by  \cite{diebold2018}. The additive version of partially egalitarian ridge or LASSO is possible,  in the sense that the solution is computable in principle. To see this, consider the simplex-constrained partially egalitarian ridge problem:
\begin{equation} \label{lasttime}
\hat{\omega} =  \min_{w}  \left (  {-\sum_{t=1}^{T} \log \left( \sum_{k=1}^{K} w_{k} f_{k,t}(y_{t})\right)}  + \lambda \sum_{k=1}^{K} \left(w_{k}-\frac{1}{\delta(w)}\right)^{2} \right) 
\end{equation} 
$$
\text{s.t. } w_{k} \in [0,1], ~ \sum_{k=1}^{K} w_{k} =1,
$$
where $\delta(\omega)$ is the number of non-zero elements in $\omega$.  Computation of the solution proceeds as follows:
 \begin{enumerate}
\item We define $\kappa$ as the number of forecasters to be included. 
\item For a particular value of $\kappa$ (among $\kappa = 1,2,3,...,K$), there are $C_{\kappa}^{K}$ possible combinations of forecasters. 
\item For the $j$th such combination ($j=1,2,...,C_{\kappa}^{K}$), we solve 
\[
L^{*}(\kappa,j)= \min_{w^{j}} \left( -\sum_{t=1}^{T} \log \left( \sum_{k=1}^{K} w_{k}^{j} f_{k,t}(y_{t})\right) + \lambda \sum_{k=1}^{K} \left(w_{k}^{j}-\frac{1}{\delta(w)}\right)^{2} \right) 
\]
\[
\text{s.t.} w_{k}^{j} \in [0,1], ~ \sum_{k=1}^{K}w_{k}^{j} = 1,
\]
where $w^{j}_{k}$ is zero if the $k$th forecaster is not selected in $j$th combination. In this case, some of weights are forced to zero, so the penalty term is reduced to
\[
\lambda \sum_{k=1}^{K} \left(w_{k}^{j}-\frac{1}{\delta(w)}\right)^{2} = \lambda \sum_{k \in \mathcal{N}} \left( w_{k}^{j} - \frac{1}{\kappa}\right)^{2},
\] 
where $\mathcal{N} = \{k: w_{k}^{j} \neq 0\}$. This is just partial egalitarian ridge for a particular set of forecasters. 

\item The solution to the original partial egalitarian ridge problem is then $\argmin_{\kappa,j} L^{*}(\kappa, j)$.

\end{enumerate}
Unfortunately, however,  the computational cost is huge, because we need to solve the penalized optimization $n_{K} =  \sum_{\kappa = 1}^{K} C_{\kappa}^{K}$ times.  For example, when $K{=}20$, $n_{K}{=}1,048,575$.  Hence partially egalitarian procedures for density mixture construction are infeasible in general.  

There is, however, one very important exception.  As $\lambda {\rightarrow} \infty$ in equation (\ref{lasttime}),  the partially egalitarian estimator converges to a direct subset averaging procedure in the spirit of \cite{Elliott2011}, which is simple to compute  and automatically imposes the simplex constraint. The subset averaging idea is trivial: At each time, rolling forward, we simply find the historically best-performing average, and use it.  A first variation is ``best $N$-Average".  At each time we  determine the historically best-performing $N$-forecast average and use it. A second variation is ``best $\mathord \le N_{max}$-Average".  At each time we determine the historically best-performing $\mathord \le N_{max}$-forecast average and use it.

Subset averaging computation time can  be substantial in principle, depending on $K$ and $N$ (or $N_{max}$). With  $K$ forecasters, finding  the best $N$-average requires computing $_{K} C_N$   simple averages and then sorting them to determine the minimum,  each period.  The per-period computational burden of best $\mathord \le N_{max}$-forecast averaging is still larger, because we now consider all subsets rather than only subsets of size $N$.  Fortunately, the relevant $K$ and $N_{max}$ are quite small in typical economic forecast combinations. In our subsequent empirical work, for example,  $N_{max} \mathord  \le 4$ appears adequate, and we have  $K \mathord  =19$.  Best $\mathord \le 4_{max}$-Average combination therefore requires evaluating and sorting just $_{19} C_4 + _{19} C_3 + _{19} C_2 + _{19} C_1 = 5035$ averages per period.

\subsection{Discussion}

It bears emphasizing that our regularized mixtures of density forecasts are not just straightforward adaptations of existing methods of combining point forecasts.  They differ in important and interesting ways.  

\begin{enumerate}

\item The objective function changes.  Things like ``forecast errors" and the ``sum of squared errors" are ill-defined in the density case.  Appropriate density forecast scoring rules must be used.  We have emphasized several, including the log score, the Brier score, and the ranked score.

\item The penalty function changes.  

\begin{enumerate}

\item When forming mixtures of density forecasts, the unit simplex constraint \textit{must} be imposed, and it has the side benefit of proving some regularization.  

\item Mixtures of density forecasts admit new regularization penalties that are intimately connected to the maintained simplex constraint, by viewing the mixture weights as a discrete probability distribution.  We introduced several such penalties, emphasizing  Kullback-Leibler distance (entropy).

\end{enumerate}

\item 
Finally (and we have not yet noted this),   it is generally  unnecessary to center regularization penalties around equal weights once the simplex constraint is imposed.  Shrinkage toward equal weights will be induced either way.  

Consider, for example, the  ridge+simplex penalty in equation (\ref{cup}), and  consider centering around equal weights, as written, vs centering around 0. There is no difference, because  
\begin{equation}
\sum_{k=1}^{K} \left (\omega_{k} - \frac{1}{K} \right)^{2} = \sum_{k=1}^{K} \omega_{k}^{2} -\frac{2}{K} \sum_{k=1}^{K} \omega_{k} + \frac{1}{K}= \sum_{k=1}^{K} \omega_{k}^{2} +\frac{1-2K}{K},
\end{equation}
where the last equality is due to the sum-to-one restriction embedded in the simplex constraint.\footnote{In fact this equivalence holds as long as all weights are centered on the same value (it does not have to be $1/K$) and the weights are constrained to sum to to a bounded real value (it does not have to be $1$).} The intuition is simply that shrinkage toward 0 is \textit{impossible} when maintaining the sum-to-one restriction, and equal weights are as close to 0 as one can get.

\end{enumerate}

\section{Monte Carlo} \label{MonteCarlo}

\begin{table}[tbp]
        \begin{center}
                \caption{Average Log Scores, DGP 1}
                \label{LOGs}
                \begin{tabular}{lrrr} 
                        \midrule 
                        \midrule 
                        Regularization group & $ L $ & \# & $\lambda^*$ \\ 
                        \midrule 
                        Simplex & -1.31 & 5.27 & NA \\ 
                        Simplex + Ridge & -1.15 & 20.00 & 2511.25 \\ 
                        Simplex + Entropy & -1.15 & 20.00 & 5.22 \\ 
                        \midrule 
                        \midrule 
                        Subset Averages & $ L $ & $\#$ & $\lambda^*$ \\ 
                        \midrule 
                        Best $N$-Average: &  &  & \\ 
                        \quad $N {=} 1$ & -2.64 & 1.00 & NA \\ 
                        \quad $N  {=}  2$ & -1.59 & 2.00 & NA \\ 
                        \quad $N  {=}  3$ & -1.37 & 3.00 & NA \\ 
                        \quad $N  {=}  4$ & -1.29 & 4.00 & NA \\ 
                        \quad $N  {=}  5$ & -1.23 & 5.00 & NA \\ 
                        \quad $N  {=}  6$ & -1.22 & 6.00 & NA \\ 
                        \quad $N  {=}  7$ & -1.21 & 7.00 & NA \\ 
                        \quad $N  {=} 8$ & -1.20 & 8.00 & NA \\ 
                        \quad $N  {=}  9$ & -1.18 & 9.00 & NA \\ 
                        \quad $N  {=}  10$ & -1.18 & 10.00 & NA \\ 
                        \quad $N  {=}  15$ & -1.16 & 15.00 & NA \\ 
                        \quad $N  {=}  20$ & -1.15 & 20.00 & NA \\ 
                        Best $ {\leq} 2$-Average & -1.61 & 2.00 & NA \\ 
                        Best $ {\leq}  3$-Average & -1.42 & 2.84 & NA \\ 
                        Best $ {\leq}  5$-Average & -1.34 & 3.63 & NA \\ 
                        Best $ {\leq}  10$-Average & -1.33 & 3.71 & NA \\ 
                        Best $ {\leq}  15$-Average & -1.33 & 3.71 & NA \\ 
                        Best $ {\leq}  20$-Average & -1.33 & 3.71 & NA \\ 
                        \midrule 
                        \midrule 
                        Comparisons & $ L $ & $\#$ & $\lambda^*$ \\ 
                        \midrule 
                        Best & -0.24 & 1 & NA \\ 
                        95th Percentile & -0.53 & 1 & NA \\ 
                        Median & -1.40 & 1 & NA \\ 
                        5th Percentile & -4.16 & 1 & NA \\ 
                        Worst & -12.19 & 1 & NA \\ 
                        \midrule 
                        Simple $K$-Average & -1.15 & 20 & NA \\ 
                        \midrule
                        \midrule
                \end{tabular} 
        \end{center}
\begin{spacing}{1} \footnotesize 
        Notes:   $L$ is the average log score,  \# is the average number of forecasters selected, $\lambda^*$ is the ex post optimal penalty parameter, and $K$ is the total number of forecasters.  We perform 10,000 Monte Carlo replications.
        \setlength{\baselineskip}{4mm}
    \end{spacing}
\end{table}%

\begin{table}[tbp]
        \begin{center}
                \caption{Average Log Scores, DGP 2}
                \label{LOGs2}
                \begin{tabular}{lcccc}
                        \midrule
                        \midrule
                        Regularization group & $ L $ & \# & $\lambda^*$ \\ 
                        \midrule 
                        Simplex & -1.29 & 4.74 & NA \\ 
                        Simplex + Ridge & -1.19 & 8.65 & 15.00 \\ 
                        Simplex + Entropy & -1.27 & 20.00 & 0.05 \\ 
                        \midrule 
                        \midrule 
                        Subset Averages & $ L $ & $\#$ & $\lambda^*$ \\ 
                        \midrule 
                        Best $N$-Average: &  &  & \\ 
                        \quad $N  {=}  1$ & -2.65 & 1.00 & NA \\ 
                        \quad $N  {=}  2$ & -1.57 & 2.00 & NA \\ 
                        \quad $N  {=}  3$ & -1.34 & 3.00 & NA \\ 
                        \quad $N  {=}  4$ & -1.26 & 4.00 & NA \\ 
                        \quad $N  {=}  5$ & -1.21 & 5.00 & NA \\ 
                        \quad $N  {=}  6$ & -1.19 & 6.00 & NA \\ 
                        \quad $N  {=}  7$ & -1.19 & 7.00 & NA \\ 
                        \quad $N  {=}  8$ & -1.18 & 8.00 & NA \\ 
                        \quad $N  {=}  9$ & -1.18 & 9.00 & NA \\ 
                        \quad $N  {=}  10$ & -1.18 & 10.00 & NA \\ 
                        \quad $N  {=}  15$ & -1.46 & 15.00 & NA \\ 
                        \quad $N  {=}  20$ & -1.64 & 20.00 & NA \\ 
                        Best $ {\leq}  2$-Average & -1.57 & 2.00 & NA \\ 
                        Best $ {\leq}  3$-Average & -1.39 & 2.83 & NA \\ 
                        Best $ {\leq}  5$-Average & -1.33 & 3.46 & NA \\ 
                        Best $ {\leq}  10$-Average & -1.33 & 3.51 & NA \\ 
                        Best $ {\leq}  15$-Average & -1.33 & 3.51 & NA \\ 
                        Best $ {\leq} 20$-Average & -1.33 & 3.51 & NA \\ 
                        \midrule 
                        \midrule 
                        Comparisons & $ L $ & $\#$ & $\lambda^*$ \\ 
                        \midrule 
                        Best & -0.28 & 1 & NA \\ 
                        95th Percentile & -0.98 & 1 & NA \\ 
                        Median & -3.79 & 1 & NA \\ 
                        5th Percentile & -32.69 & 1 & NA \\ 
                        Worst & -182.42 & 1 & NA \\ 
                        \midrule 
                        Simple $K$-Average & -1.64 & 20 & NA \\ 
                        \midrule
                        \midrule
                \end{tabular}%
        \end{center}
    \begin{spacing}{1} \footnotesize 
        Notes:   $L$ is the average log score,  \# is the average number of forecasters selected,  $\lambda^*$ is the ex post optimal penalty parameter, and $K$ is the total number of forecasters.   We perform 10,000 Monte Carlo replications. 
        \setlength{\baselineskip}{4mm}
    \end{spacing}
\end{table}%

\begin{figure}[tp]
        \begin{center}
                \floatpagestyle{empty}
        \caption{Monte Carlo Estimates of Expected Mixture  Performance vs Penalty Strength}
        \label{onemore}
        \includegraphics[trim= 20mm 20mm 0mm 20mm, clip, scale=.15]{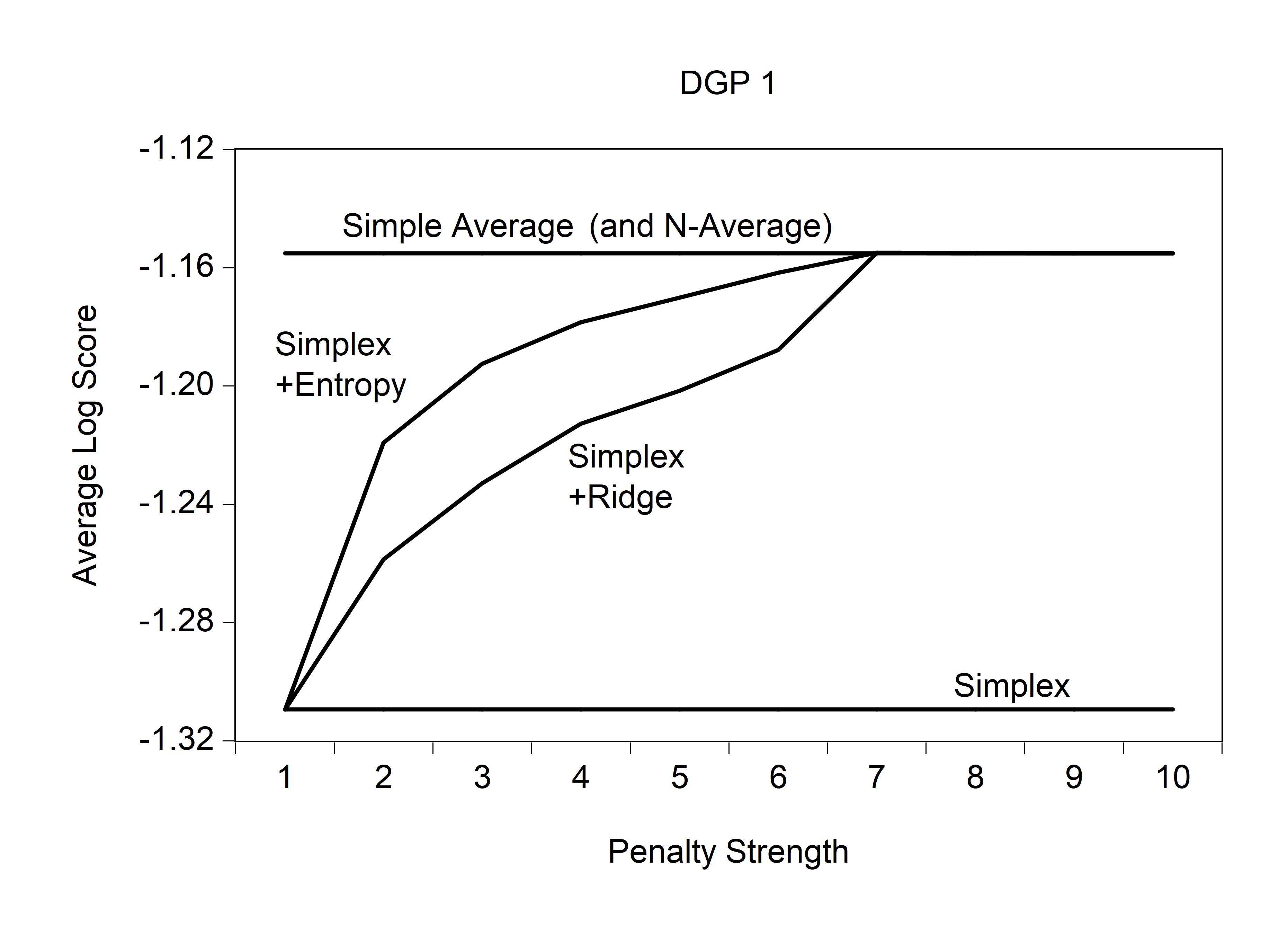}
        \includegraphics[trim= 20mm 0mm 0mm 0mm, clip, scale=.15]{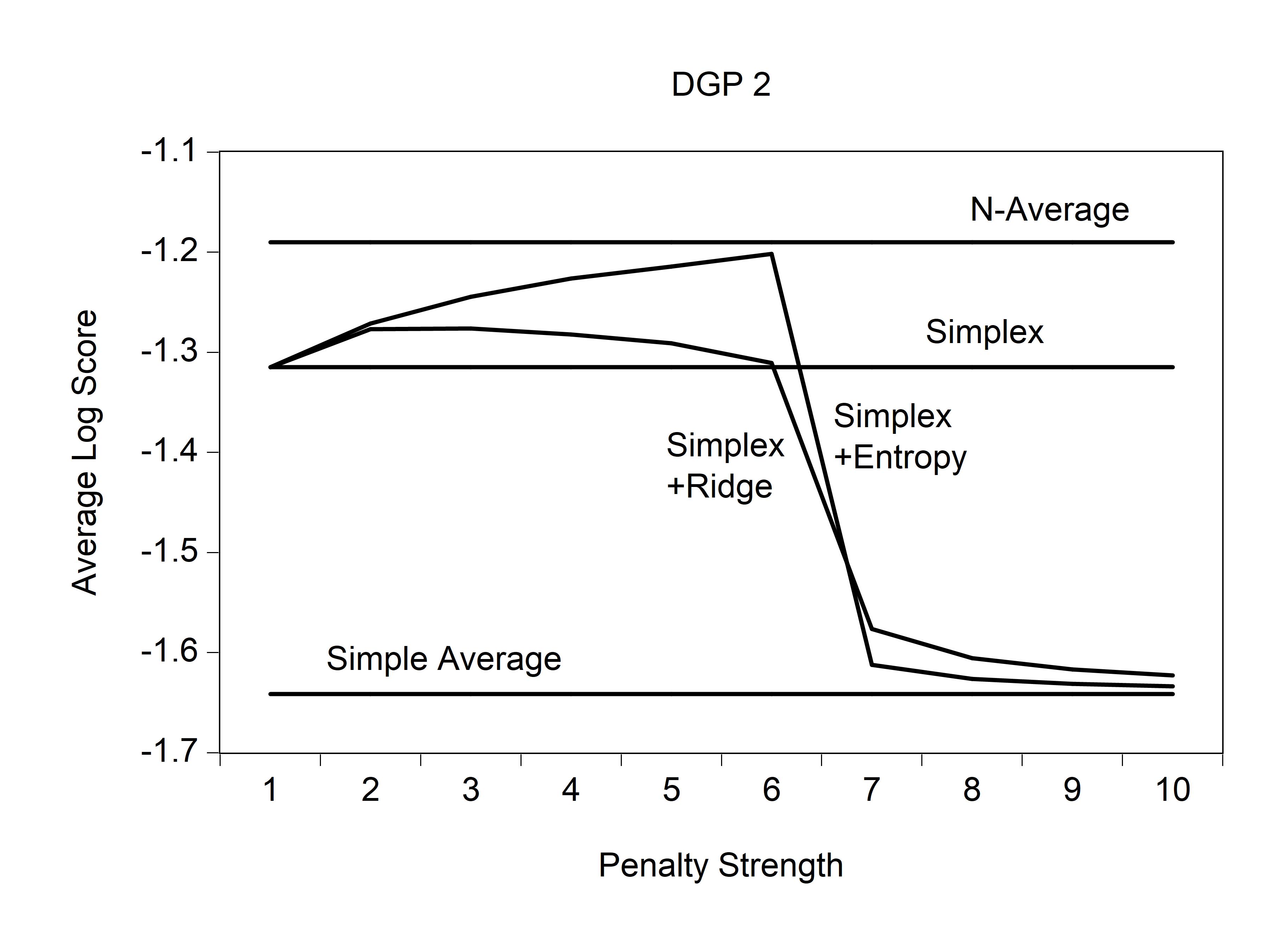}
\end{center}
\begin{spacing}{1}
        Notes: We perform 10,000 Monte Carlo replications.
\end{spacing}
\end{figure}

We now explore the potential of our regularized mixture estimators via a small Monte Carlo analysis.  The data-generating process (DGP), which we assume to be known by the forecasters, is:
\begin{equation}
\begin{split}
y_{t} &= x_{t} + \sigma_{y} e_{t}, \quad e_{t} \sim {\rm iid \,}N(0,1) \\
x_{t} &= \phi_{x} x_{t-1} + \sigma_{x} v_{t}, \quad v_{t} \sim  {\rm iid \,}N(0,1), \\
\end{split}
\end{equation}
where $e$ and $v$ are orthogonal at all leads and lags.  $y$ is  the variable to be forecast, and   $x_{t}$ can be interpreted as the long-run component of $y_{t}$. Individual forecasters receive heterogeneous independent noisy signals about $x_{t}$.  For forecaster $k$ we have 
\begin{equation}
z_{kt} = x_{t} + \sigma_{zk} \eta_{kt}, \quad \eta_{kt} \sim  {\rm iid \,} N(0,1),
\end{equation}
 where $\eta_{k}$ and $\eta_{k'}$  are orthogonal at all leads and lags for all forecasters $k$ and $k'$. Assume that  forecasters have a strong  belief that  the 1-step-ahead predictive density is Gaussian with variance $\sigma_{y}^{2}$, but that they don't know its mean, and that forecaster  $k$ therefore uses $z_{kt}$, resulting in the  predictive density
\begin{equation}
p_{kt}(y_{t+1}) = N(\phi z_{kt}, \sigma_{y}^{2}).
\end{equation}
Note that in this environment, forecasters' predictive densities differ only by their locations (means).

We consider two parameterizations: 
\begin{enumerate}
\item DGP 1:  $\sigma_{zk} {=}1$ for all $k$
\item DGP 2: $\sigma_{zk} {=} 1$ for $k=1,2,...,\frac{K}{2}$ and $\sigma_{zk}{=}5$ for $k=\frac{K}{2}{+}1, ..., K$, 
\end{enumerate}
where each DGP has common parameters $\phi_{x}{=} 0.9$, $\sigma_{x} {=}1$, $\sigma_{y} {=} 0.5$.  The two DGPs  differ only by the quality of the signals that forecasters receive. Under DGP 1 the simple average should be preferred, because all signals are of the same quality, while under  DGP 2 the linear opinion rule should be preferred (at least asymptotically, so that estimation error vanishes), giving more weight to forecasters  $k=1,2,...,\frac{K}{2}$, who receive better signals.

To cohere with our subsequent  empirical work, we explore $K{=}T{=}20$. We generate data, estimate mixture weights, generate 1-step-ahead mixture densities, and evaluate them using the log score  objective. We repeat this 10,000 times and compute the average LPS for several methods:  

\begin{enumerate}
        
\item Simple Average 

\item Simplex (equation  (\ref{AG}))

\item Simplex+Ridge (equation (\ref{cup}))

\item  Simplex+Entropy (equation (\ref{log2}))

\item Subset Averaging  (equation (\ref{lasttime}) with  $\lambda {\rightarrow} \infty$).

\end{enumerate}
For each of simplex+ridge and simplex+entropy, we explore  20 penalization strengths. {For simplex+ridge, we choose 10  equispaced  points in [1e-15,10] and 10 equispaced points in [15,10000]. For simplex+entropy we choose 10 equispaced points in [1e-15,0.2] and 10 equispaced points in [0.3,20].}

Numerical results appear in Tables \ref{LOGs} and \ref{LOGs2}, in which  we present the the optimized average log score for each method under DGPs 1 and 2, respectively. Graphical results appear in Figure \ref{onemore}, in which we show how the optimized score varies with regularization penalty strength  under DGPs 1 and 2, respectively.  Under DGP 1, simple averaging performs well, and unregularized simplex performs poorly, as expected.  As the strength of shrinkage gets heavier, the performance of both simplex+entropy and simplex+ridge improves monotonically until they perform as well as the simple average (full shrinkage). In addition, the performance of simplex+entropy improves more quickly than that of simplex+ridge as shrinkage strength increases and dominates throughout.  Finally, subset averaging performs admirably under DGP 1, and as expected the optimal ``subset" includes all forecasters.

Under  DGP 2, simplex is expected to perform well, and simple averaging is expected to perform poorly.  Simplex does indeed outperform simple averaging.  Moreover, both simplex+ridge and simplex+entropy behave as expected. For little shrinkage (toward the left), their performance is similar to that of simplex, and for  heavy shrinkage (toward the right), their performance is similar to that of the simple average. In between, for moderate amounts of shrinkage,  they outperform simplex. In that region, regularized simplex improves on unregularized simplex, because  the large unregularized simplex estimation error makes it likely that some relevant forecasters are dropped from the pool, and regularization brings them back.  Importantly, subset averaging continues to perform admirably under DGP 2, but now the optimal average involves only 10 or so forecasters, as expected.

It is important to note that the performance documented in  Tables \ref{LOGs} and \ref{LOGs2},  and in Figure \ref{onemore}, is almost surely not achievable in practice, because it requires ex post omniscience (use of the ex post optimal penalty parameter for the regularized estimators, and use of the ex post optimal $N$ for the $N$-averages.)  Nevertheless the results are informative, because they document what can be achieved \textit{in principle}, even if not in practice.  Practical performance is an empirical matter, to which we now turn, in a detailed application to density forecasts of Eurozone inflation and real interest rates.

\section{Eurozone Inflation and Real Interest Rate Forecasts} \label{empirical}

Here we use our methods to construct regularized mixtures of density forecasts for Eurozone inflation and real interest rates. Expected inflation is a key driver of the bond market via its direct impact on nominal interest rates.   Expected  inflation may also negatively impact real growth, and hence the stock market, insofar as it  ``puts sand in the Walrasian gears", as classically emphasized by  \cite{inflation}.  High inflation, moreover, also tends to be \textit{volatile} inflation \citep{friedman}, which adds additional sand.\footnote{See also \cite{chen1986economic}.}  Expected inflation is also a key part of the ex ante real interest rate, which  in turn   is a key guide to intertemporal allocation and a key link between macroeconomic fundamentals and financial markets. From a variety of angles, then, inflation forecasts are central to financial markets, the macroeconomy, and the interface.

\subsection{Data}

\begin{figure}[tp]
        \begin{center}
                \caption{Individual and Average Density Forecasts, Eurozone Inflation, 2004Q4 (left) and 2018Q4 (right)} 
                \label{fig1}
                \includegraphics[trim= 0mm 2mm 0mm 0mm, clip, width=3.1in]{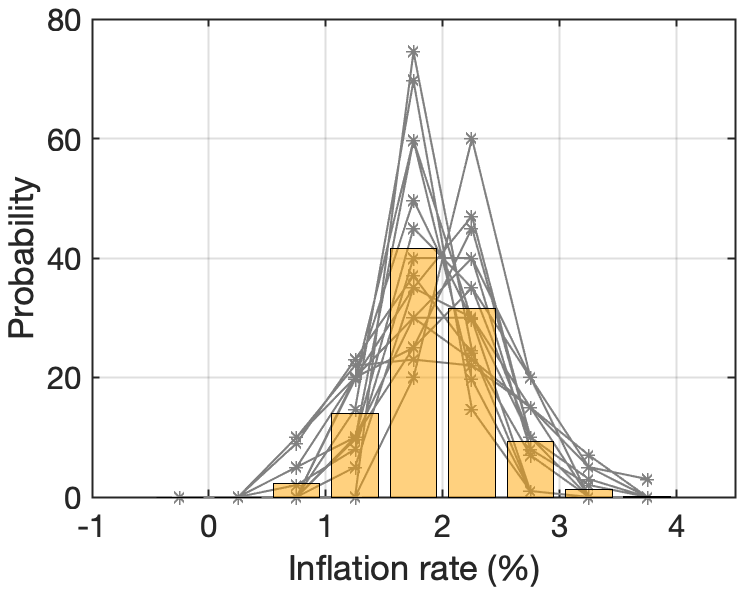} ~
                \includegraphics[trim= 0mm 2mm 0mm 0mm, clip, width=3.1in]{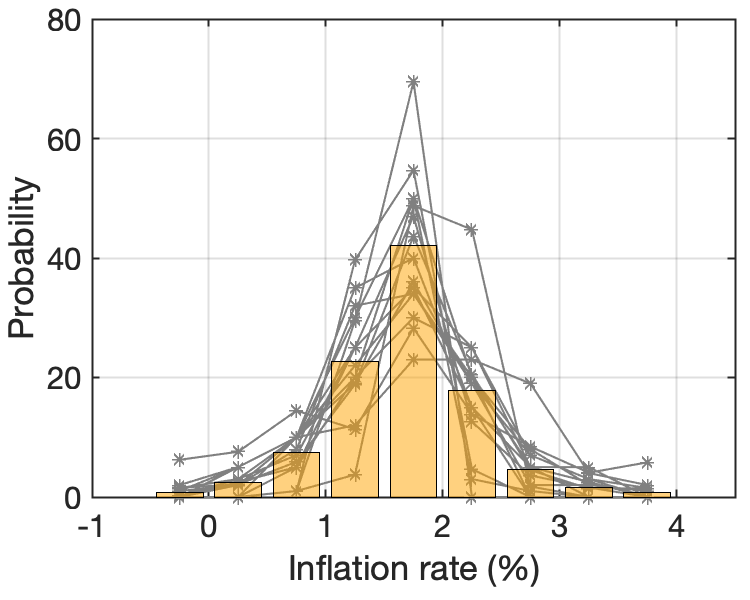}
        \end{center}
        \begin{spacing}{1}  \footnotesize 
                Notes:  We show the individual survey forecasts in gray (as frequency polygons), and the average forecast in orange (as a histogram). 
        \end{spacing}
\end{figure}

Following the pathbreaking work of \cite{conflitti2015}, we study  inflation density forecasts from the European Central Bank Survey of Professional Forecasters (ECB-SPF), which has been undertaken since 1999.  Participants are surveyed quarterly, in January, April, July, and October.\footnote{See \url{https://www.ecb.europa.eu/stats/ecb_surveys/survey_of_professional_forecasters/html/index.en.html}.} Our forecast sample contains 83 quarterly surveys, starting in 1999Q1 and ending in 2019Q3.  

As an entr\'{e}e into the data, in Figure \ref{fig1}  we show all forecasts expressed as frequency polygons, and the simple average forecast expressed as a histogram,  for two illustrative surveys (2004Q4, 2018Q4).  Substantial differences are apparent at the two survey dates. The simple average forecast in 2004Q4, for example, puts 2.3\% probability on the event that the inflation rate is less than 1\%, whereas in 2018Q4 it puts 10.5\% probability on the same event. Continuing, in the top panel of Figure \ref{fig2}  we show the complete time series of simple average forecasts.  Again, large movements are evident over time, in both location and scale.

The precise Euro-area  inflation forecast target is the percentage change in the Harmonised [sic] Index of Consumer Prices (HICP),  for the year following the forecast.\footnote{Eurostat, Harmonized Index of Consumer Prices: All Items for Euro area (19 countries) [CP0000EZ19M086NEST], Retrieved from FRED, Federal Reserve Bank of St. Louis; \url{https://fred.stlouisfed.org/series/CP0000EZ19M086NEST}.}   For example, when the survey was conducted in October 2017 (2017Q4), HICP inflation data were available up to September 2017, so the  2017Q4 survey asks for a forecast for the year from October 2017 through September 2018. Our realization sample, matched to our forecast sample, contains 83 quarterly observations,  starting in December 1999 and ending in June 2020.

We will soon obtain mixture densities using the log score objective and several regularizations, including simplex, simplex+ridge, simplex+entropy, and subset averaging. Before proceeding to empirical results, however, we address several issues. 

\subsubsection{Survey Entry and Exit}

First, forecasters can enter and exit the survey pool. There are 103 unique forecasters between 1999Q1 and 2019Q4, and no forecaster appears in the pool  continuously. Following \cite{genre2013}, we proceed by first excluding forecasters who miss more than four consecutive surveys, which leaves 18 forecasters.  Then we  interpolate the remaining gaps based on historical performance.\footnote{More precisely, we fill in the gaps in the first survey (t=1, 1999Q1) with the average of non-missing forecasts from all other available forecasters. Then we calculate the ranked score for each forecaster and divide them into five mutually exclusive groups based on the score, and move to the second survey. At each of the following rounds ($t = 2, 3, ..., T$), we set the missing observations of a particular forecaster to the average of non-missing forecasts from her group, and then using the full set of forecasts we re-calculate ranked scores and update the group structure for use in the next round.}

\subsubsection{Time-Varying Bin Definitions}

Second, outcome bin definitions vary over time.  Although bin definitions have been stable for mid-range ``standard" inflation values, extreme tail bins have become finer over time, as realizations fell in the tails. For example, for high inflation,  there was originally a ${>} 3.5$ bin, but it  was eventually split into 3.5-4 and ${>} 4$ bins.\footnote{During our sample period the number of bins started at 9,  peaked at 14 during the Great recession, and eventually dropped to 12.}  We proceed by merging extreme tail bins sufficiently to produce 11 bin definitions, fixed for the entire sample: $(-\infty, -0.5]$, $(-0.5, 0]$, $(0, 0.5]$, ...,   $(3.5,4]$, $(4, \infty]$. 

\subsubsection{Zero-Probability Realizations}

Finally, complications can arise with the log-score objective. Consider, for example, the survey  forecast: 
\begin{equation}
y \in 
\begin{cases}
(-\infty, 1.5] ~~w.p.=0\\
(1.5, 2.0] ~~w.p.=.3\\
(2.0, 2.5] ~~ w.p.=.5\\
(2.5, 3.0] ~~w.p.=.2\\
(3.0, \infty] ~~w.p.=0.\\
\end{cases}
\end{equation}
The zero probabilities assigned to the leftmost and rightmost bins obviously create a problem (infinite loss) for the log-score objective, due to its use of logs, if a realization occurs that was assigned zero probability.

 Zero-probability realizations rarely, but occasionally,  appear in our data.  Sometimes they occur in edge bins (e.g.,  $(4, \infty]$), because forecasters sometimes fail to put positive probability on those bins. In addition to the edge-bin phenomenon, some forecasters' histograms are simply too sharp, and they sometimes put zero probability on an interior bin that eventually contains the realization.

 One can address  the log score ``zero problem" by requiring the survey bin into which the realization falls to have been assigned at least some small probability, say  1\%.  We achieve this by assigning 1\% probability to the bin containing the realization if it had originally been assigned 0, where the 1\% is taken in equal shares from the bins originally assigned non-zero probability.\footnote{One could of course switch to another objective, but the log score objective is simple and deservedly popular, which is why we have used it throughout this paper as a leading case for both our theory and Monte Carlo.  We will continue to use it for our empirical work, where it is also deservedly popular, despite the zero problem.}
 
\begin{table}[t] 
        \begin{center} 
                \caption{Log Scores for Eurozone Inflation} 
                \label{mainresults2b} 
                \scalebox{0.8}{
                        \begin{tabular}{lrrr} 
                                \toprule 
                                \toprule 
                                Regularized Mixtures & $ L $ & \# &  \\ 
                                \midrule 
                                Simplex & -1.88 & 3.52 &  \\ 
                                Simplex+Ridge & -1.86 & 4.99 &  \\ 
                                Simplex+Entropy & -1.87 & 19 &  \\ 
                                Best $4$-Average: &-1.87  &4  & \\ 
                                 Best ${\leq}4$-Average & -1.90 & 2.24 &  \\ 
                                \midrule 
                                \midrule 
                                ECB/SPF Comparisons & $ L $ & $\#$ & \\ 
                                \midrule 
                                Best & -2.02 & 1 &  \\ 
                                90\% & -2.04 & 1 &  \\ 
                                70\% & -2.13 & 1 &  \\ 
                                Median & -2.17 & 1 &  \\ 
                                Worst & -2.56 & 1 &  \\ 
                                \midrule 
                                Simple Average & -1.98 & 18 &  \\ 
                                \bottomrule 
                                \bottomrule 
                        \end{tabular} 
                }
        \end{center} \normalsize  \footnotesize 
        \begin{spacing}{1}  Notes:  We show log scores for 1-year-ahead Eurozone inflation density forecasts, made quarterly, using a 20-quarter rolling estimation window. The burn-in sample is 1999Q1-2000Q4, and the forecast evaluation sample is 2001Q1-2019Q3 (75 quarters). There are 18  ECB-SPF density forecasters in the pool, plus  a 19th forecaster whose predictive density is constant and uniform, for a total of 19 forecasters. $L$ is the log score, and \# is the average number of forecasters selected. Results for Simplex+Ridge and Simplex+Entropy are based on ex post optimal penalty parameters. See text for details. 
                
        \end{spacing}
\end{table}

\subsection{Empirical Results}

\begin{figure}[t]  
        \begin{center}
                \caption{Density Forecast Mixtures Over Time, Eurozone Inflation} 
                \label{fig2}
                \bigskip
                \includegraphics[trim= 0mm 0mm 0mm 0mm, clip, scale=0.25]{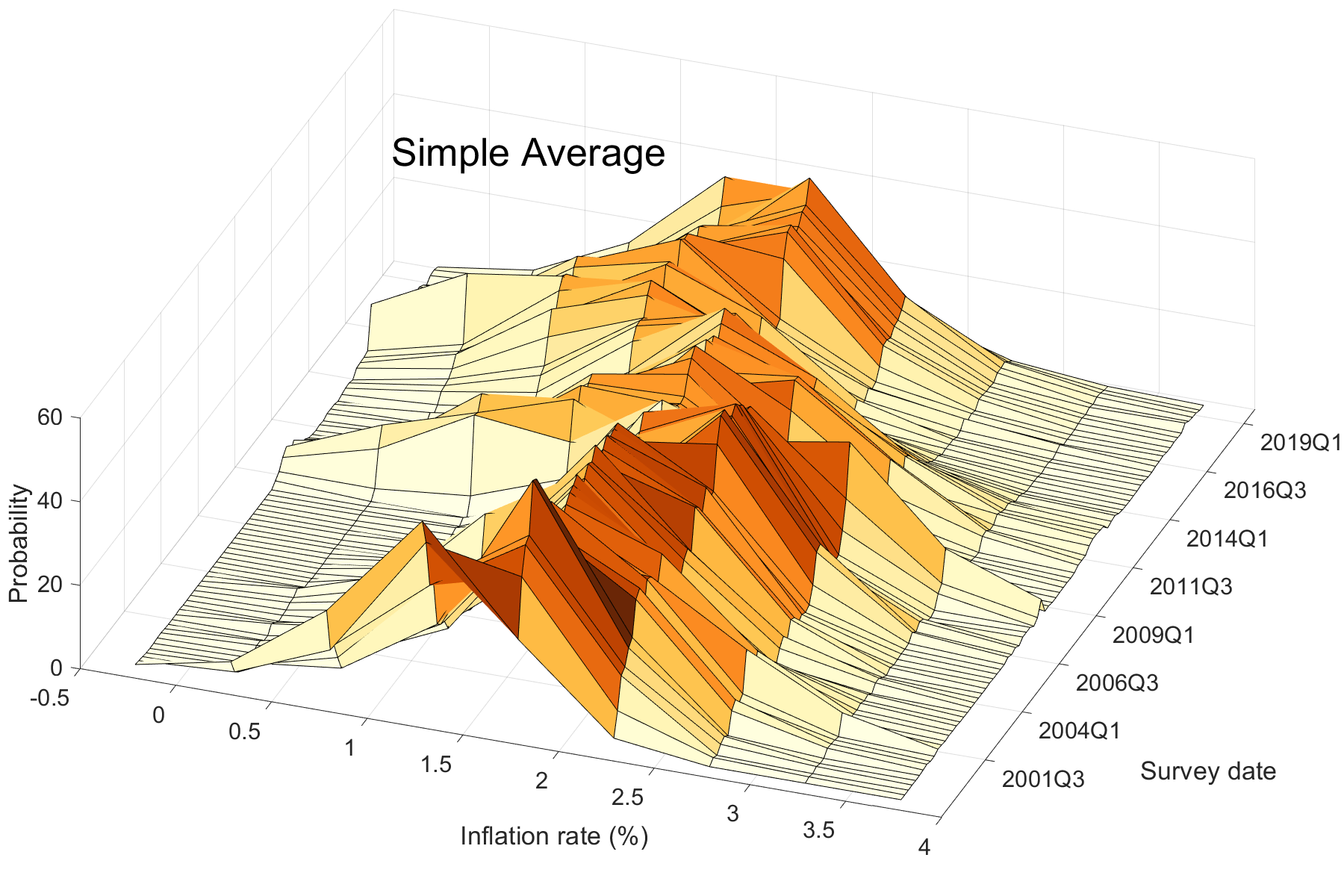}\\
                \includegraphics[trim= 0mm 0mm 0mm 0mm, clip, scale=0.250]{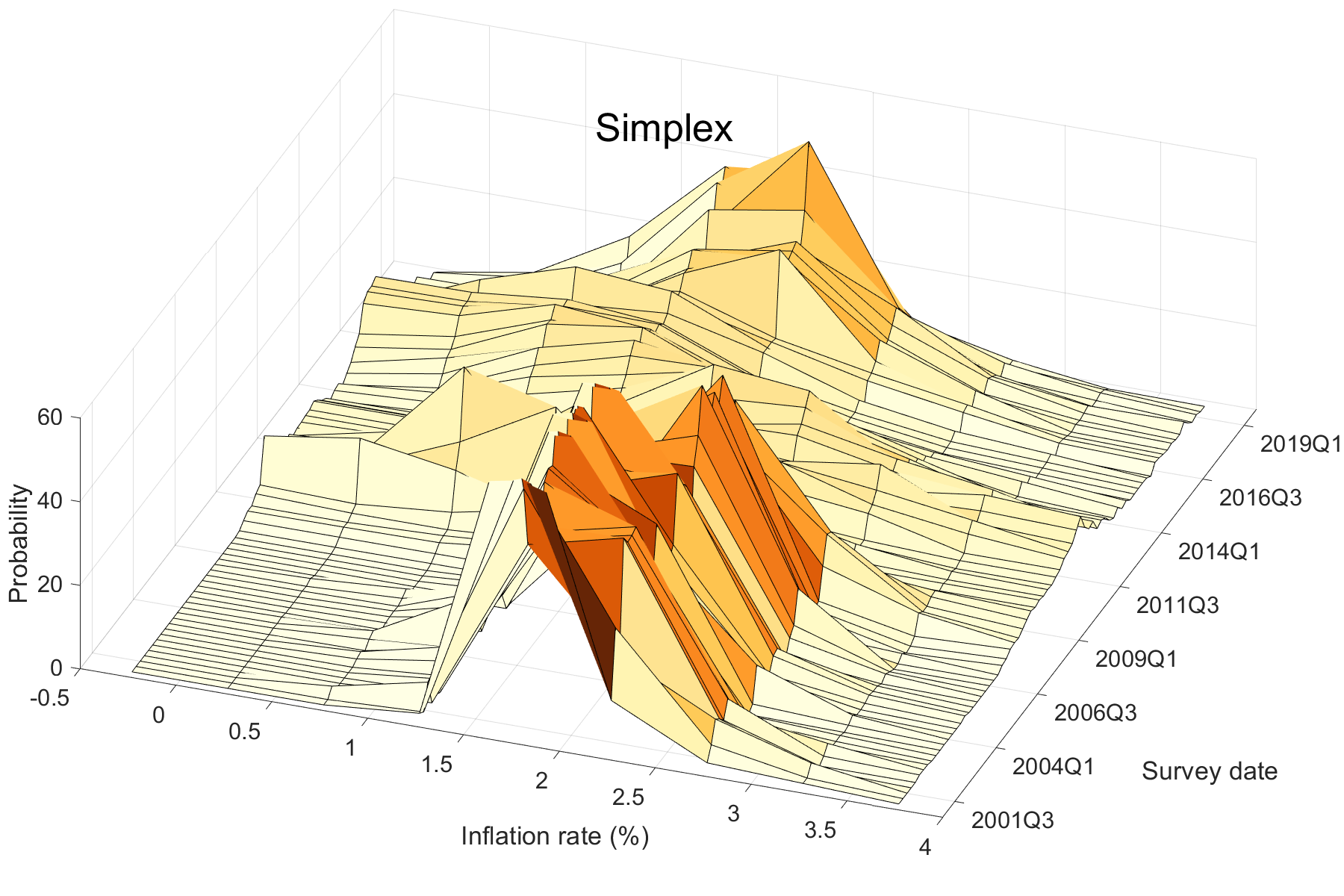}~
                \includegraphics[trim= 0mm 0mm 0mm 0mm, clip, scale=0.250]{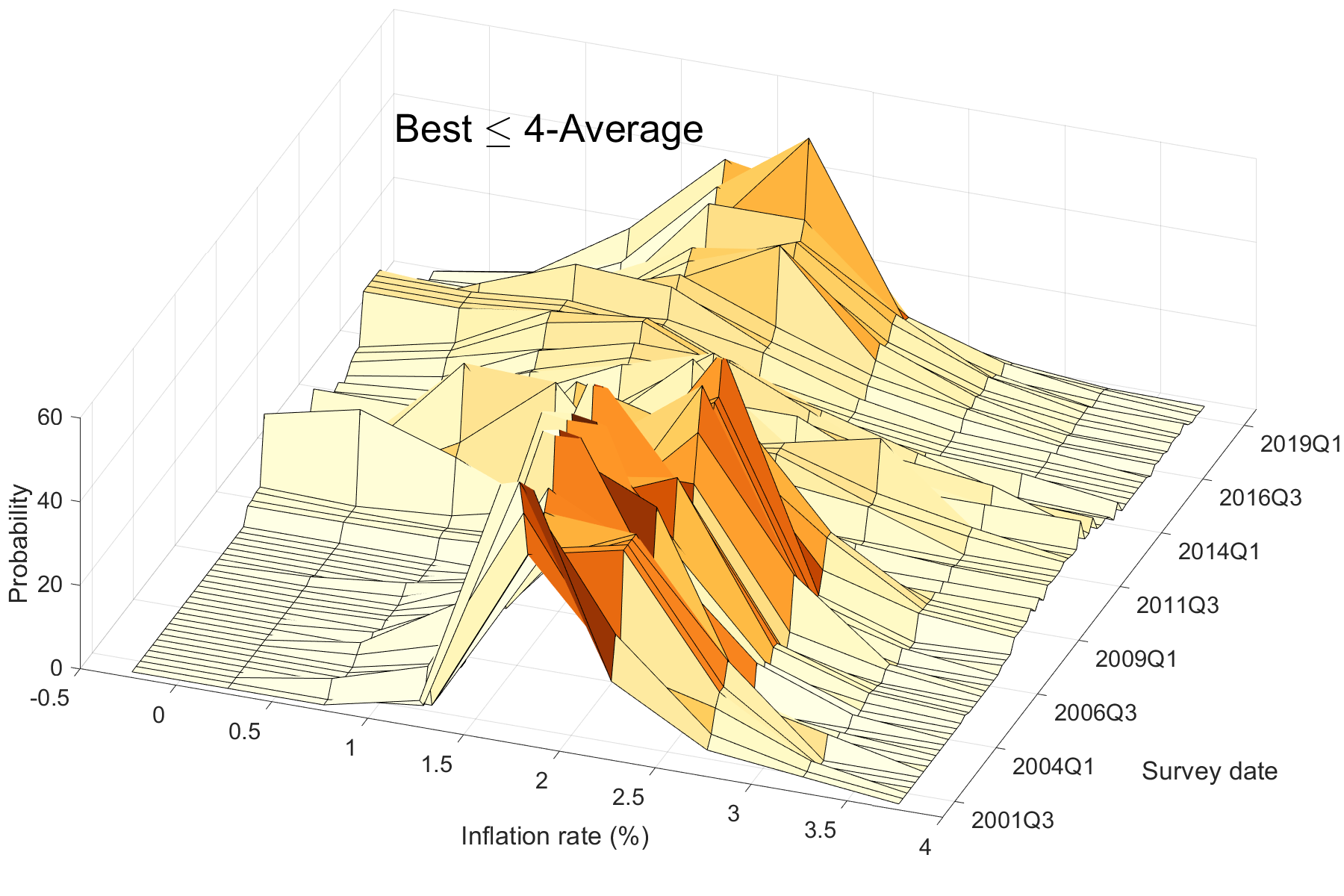}
        \end{center}
        \begin{spacing}{1} \footnotesize 
                Notes: We show density forecast mixtures expressed as frequency polygons.  The forecasts are quarterly, from 1999Q1 to 2019Q3.
        \end{spacing}    
\end{figure}

There are  18  ECB-SPF density forecasters in the pool.  We also include a fictitious 19th forecaster whose predictive density is constant and uniform, in rough parallel to including a constant  in point forecast combining regressions, for a total of 19 forecasters.  Doing so appears  desirable \textit{a priori} in the spirit of \cite{granger1984}.\footnote{Moreover, it  constrains the mixture density to put positive probability on each histogram bin as long as the uniform forecaster gets a non-zero mixture weight, in which case the earlier-discussed log score ``zero problem" vanishes.}

Results appear in Table \ref{mainresults2b}. Strikingly, each regularized mixture outperforms each ECB/SPF individual forecaster (even the ex post \textit{best} forecaster). To get a feel for the size of the improvement, note that the log score of the Best ${\leq}4$-Average, for example, is approximately 15\% better than that of the median individual forecaster, and 7\% better than that of the ex post best individual forecaster. Each regularized mixture also outperforms the simple average, which in turn outperforms the ECB/SPF forecasts.

\begin{figure}[t]
        \begin{center}
                \caption{Differences Between Regularized Mixtures and the Simple Average Mixture, Eurozone Inflation}
                \label{fig:diff_heatmap_avg}
                \includegraphics[trim= 0mm 0mm 0mm 0mm, clip, scale=0.35]{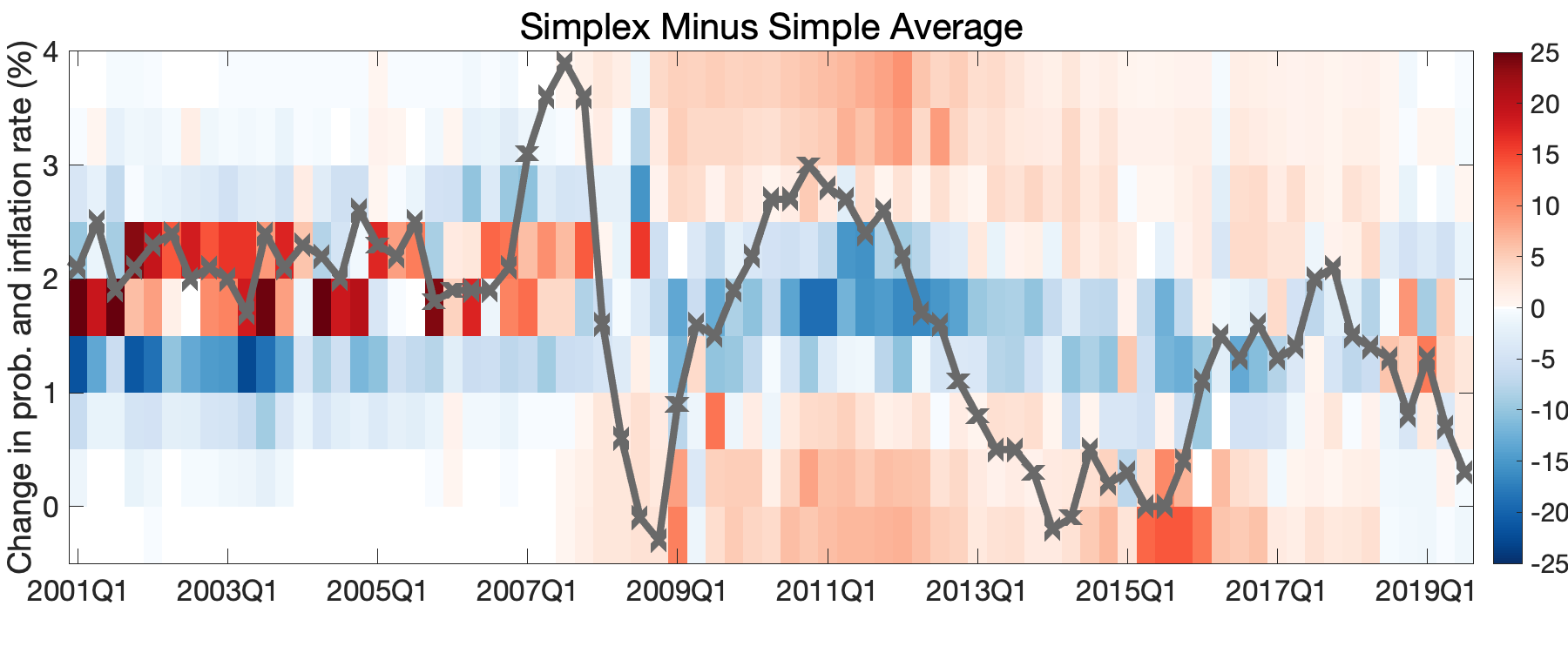}\\
                \bigskip 
                \includegraphics[trim= 0mm 0mm 0mm 0mm, clip, scale=0.35]{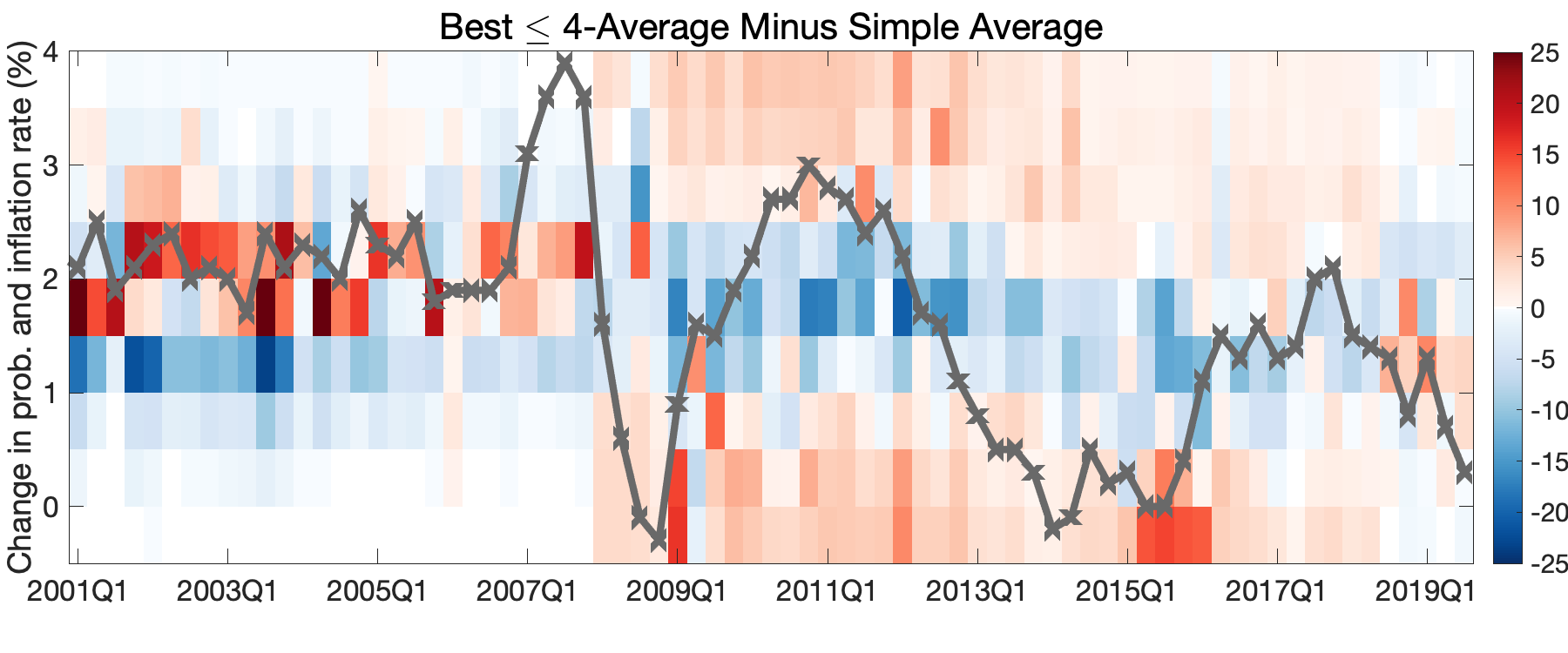} 
        \end{center}
        \begin{spacing}{1} \footnotesize   Notes:  We show heat maps of differences between a regularized mixture (Simplex or Best ${\leq}4$-Average) and the simple average mixture.  Red shadings indicate bin probability increases in the Simplex or Best ${\leq}4$-Average regularized mixture, and blue  shadings indicate bin probability decreases.  We also superimpose the realized Eurozone inflation rate.
        \end{spacing}
\end{figure}

\begin{figure}[t]  
        \begin{center}
                \caption{\textit{PIT} Histograms, Eurozone Inflation}
                \label{aaabbb}
                \includegraphics[height=3.4in]{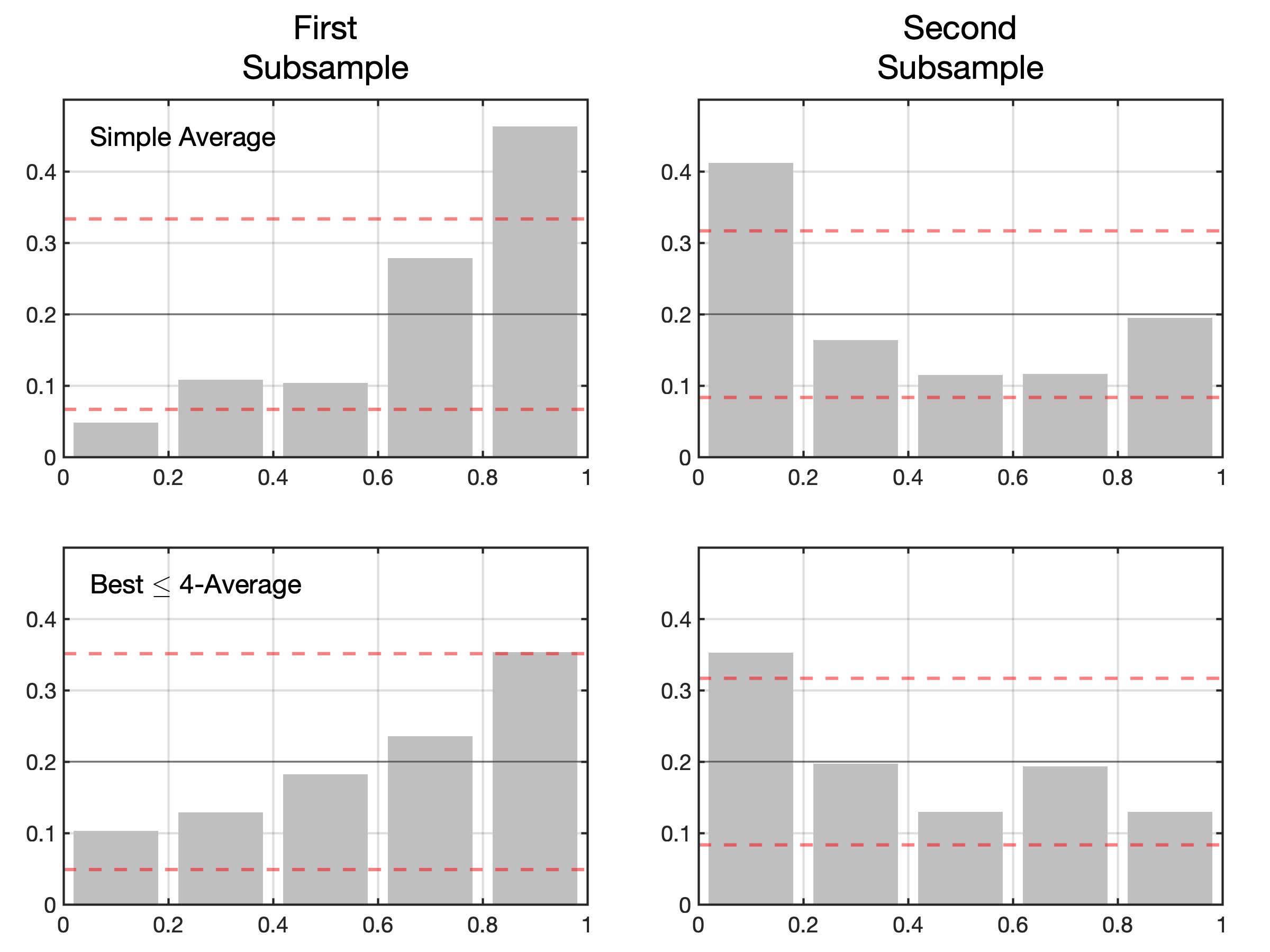}\\
        \end{center}
        \begin{spacing}{1} \footnotesize   Notes:  We show $PIT$ histograms for Simple Average and  Best ${\leq}4$-Average mixtures. The first subsample ends in 2007Q4, and the second subsample begins in 2008Q1. We show pointwise binomial confidence bands in red under  $PIT \sim  iid \, U(0,1)$. 
        \end{spacing}
\end{figure}

\begin{figure}[tb]
        \begin{center}
                \caption{Density Forecast Mixtures Over Time, Eurozone Real Interest Rate} 
                \label{realint3D}
                \includegraphics[trim= 0mm 0mm 0mm 0mm, clip, scale=0.250]{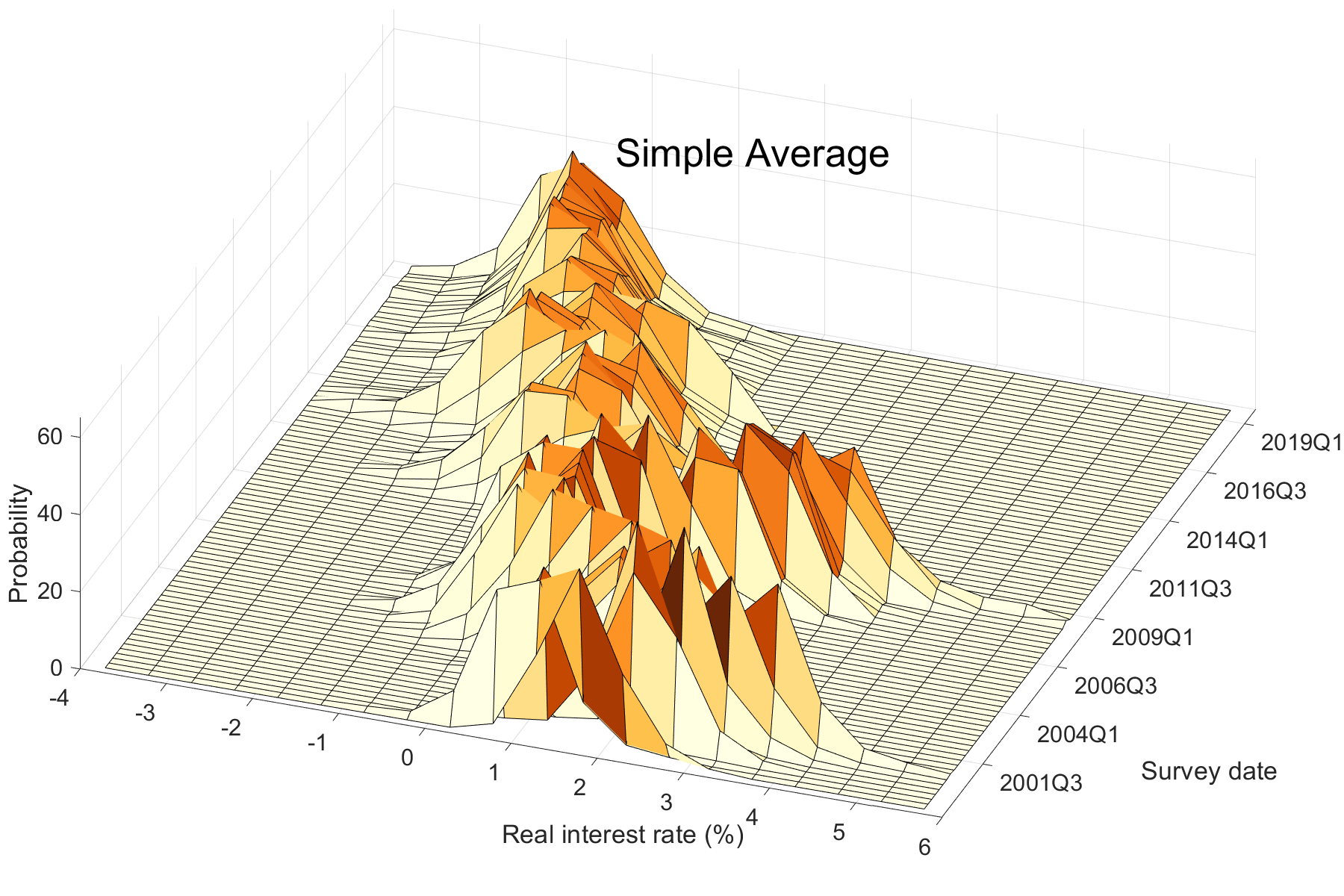} 
                \includegraphics[trim= 0mm 0mm 0mm 0mm, clip, scale=0.250]{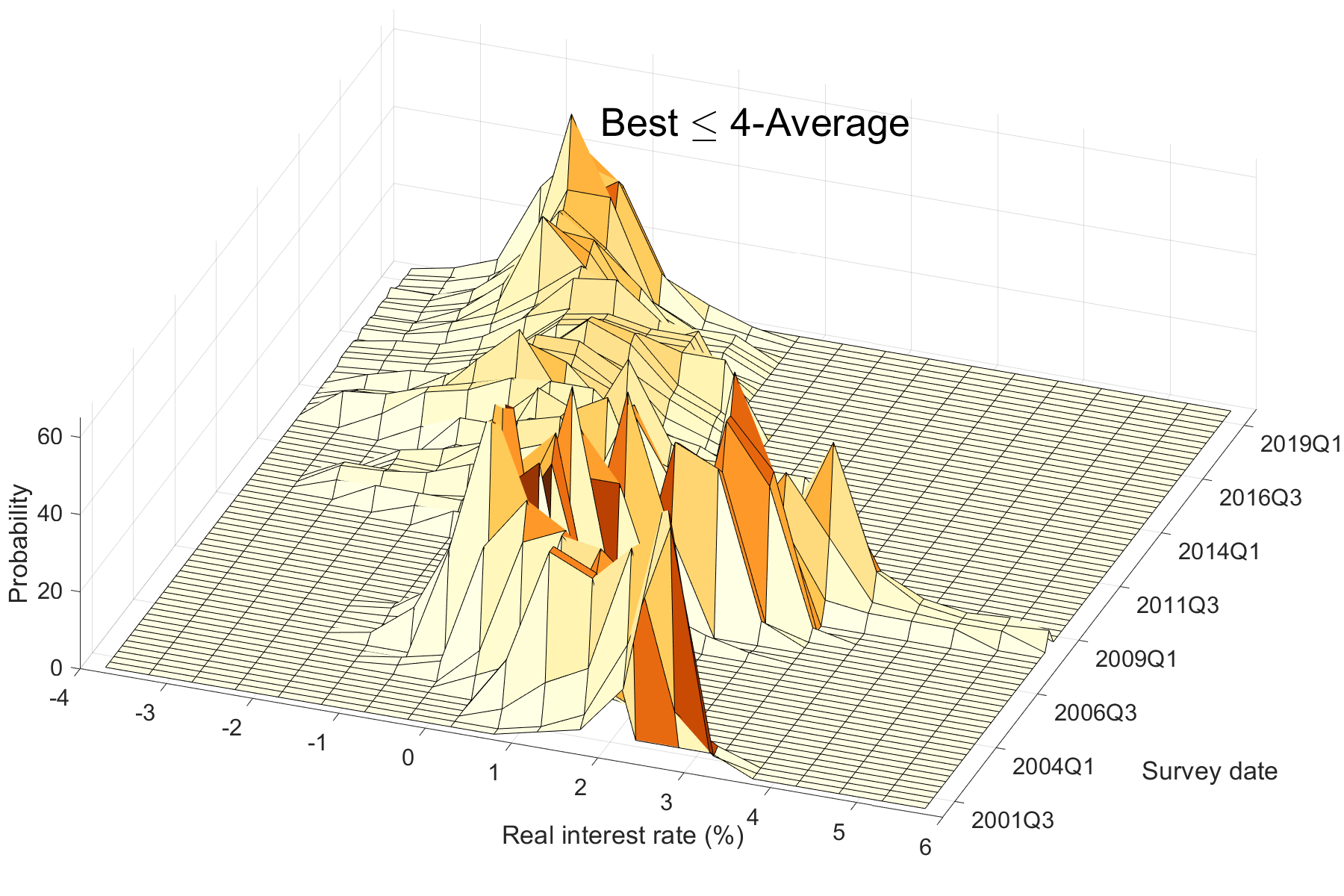}
        \end{center}
        \begin{spacing}{1} \footnotesize 
                Notes: We show density forecast mixtures expressed as frequency polygons.  The forecasts are quarterly, from 1999Q1 to 2019Q3.
        \end{spacing}    
\end{figure}

\begin{figure}[tb]
        \begin{center}
                \caption{Difference Between the Best ${\leq}4$-Average Mixture and the Simple Average Mixture, Eurozone Real Interest Rate}
                \label{fig:diff_heatmap_avg realint}
                \includegraphics[trim= 0mm 0mm 0mm 0mm, clip, scale=0.4]{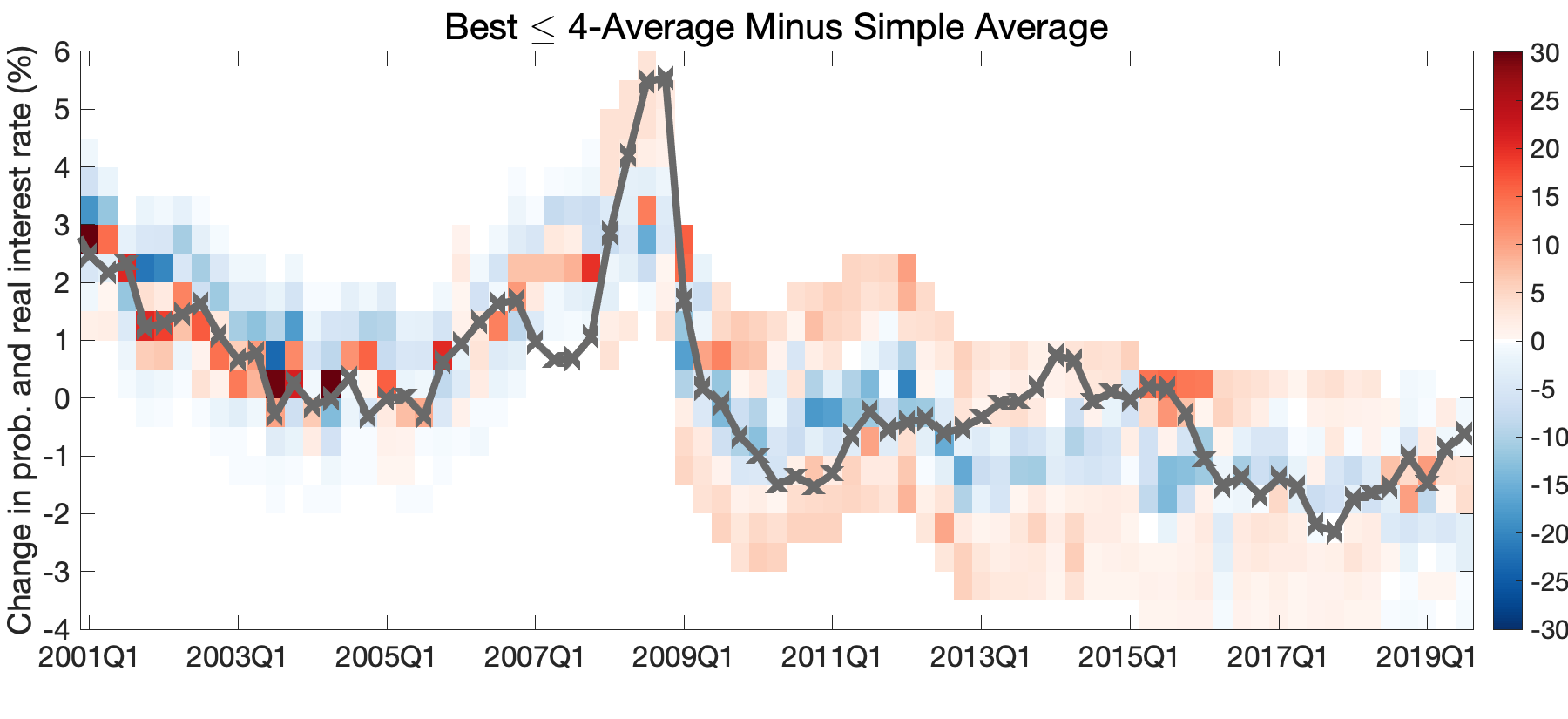} 
        \end{center}
        \begin{spacing}{1}  \footnotesize  Notes:  We show a heat map of the difference between the Best ${\leq}4$-Average mixture and the Simple Average mixture.  Red shadings indicate probability increases in the Best ${\leq}4$-Average mixture, and blue  shadings indicate probability decreases.   We also superimpose the realized Eurozone real interest rate.
        \end{spacing}
\end{figure}

Table \ref{mainresults2b} also reveals that the average number of forecasters selected after regularization is always  small, regardless of the regularization method.\footnote{Simplex+Entropy selects all 19 forecasters, but Simplex+Entropy \textit{must} select  all 19 forecasters, because $\log(\omega_{k}) {\rightarrow} \infty$ as $\omega_{k} {\rightarrow} 0$.  All regularizations capable of selecting only a few forecasters do in fact  select only a few.} Simultaneously, both the log scores in Table  \ref{mainresults2b} and the graphs in the bottom two panels of Figure \ref{fig2} reveal that the Simplex and  Best Average regularized mixtures are almost identical, suggesting that the Simplex solution is effectively dropping all but a few forecasts and simply averaging the survivors, producing something very close to a Best 4-Average.  

The good performance of both Simplex and Best Average is particularly noteworthy insofar as they do not require tuning.\footnote{Strictly speaking,  Best Average procedures require some slight tuning -- a choice of $N$ -- although we are comfortable with simply always adopting  $N{=4}$.}  That is, quite remarkably, the Simplex and Best Average regularizations perform as well as those requiring choice of tuning parameters (Simplex+Ridge and Simplex+Entropy), despite the fact that we evaluate the latter in Table \ref{mainresults2b} using ex post optimal tuning parameters, which is not feasible in real time. 

 Figure \ref{fig2}  merits additional examination.  If its middle and bottom panels reveal that the Simplex and Best Average regularized mixtures are nearly identical, a comparison of those panels with the top panel also reveals that (1) Simplex / Best Average regularization is nevertheless \textit{very} different from a simple average, and (2) the effects  of Simplex / Best Average regularization  differ strikingly before and after the onset of the Great Recession.   Before the onset of the Great Recession, Simplex / Best Average regularization moves probability mass upward toward higher inflation relative to simple averaging, particularly from the 1.0\%-1.5\% range to the 1.5\%-2.5\% range, mostly adjusting density forecast location and symmetry.  After that, however, Simplex / Best Average regularization spreads probability mass from the center  into both tails of the distribution, from the 1.0\%-2.5\% range outward to below 0.5\%  and above 3.0\%, mostly adjusting density forecast dispersion and kurtosis. The regularization effects, and their structural shift at the onset of the Great Recession, are revealed even more clearly in the heatmaps shown in Figure  \ref{fig:diff_heatmap_avg}.

It is informative to examine and compare  probability integral transforms ($PIT$s) for various mixtures. \cite{diebold1998} consider the continuous case, in which the $PIT$ is defined as 
$PIT_{t} = \int_{-\infty}^{y_t} \, p_{t}(u) du$, and show that correct conditional calibration of density forecasts implies that  $PIT \sim  iid \, U(0,1)$.  \cite{czado2009} extend the evaluation framework to the discrete case and show that the result still holds for an appropriate discrete $PIT$ definition. To assess uniformity, and any patterns in deviations from uniformity,  in Figure \ref{aaabbb}  we show  histograms of the \cite{czado2009} discrete $PIT$ for the Simple Average and Best ${\leq} 4$-Average mixtures.\footnote{There is no need to show the Simplex $PIT$ because the Simplex and  Best ${\leq} 4$-Average mixtures are almost identical, as discussed earlier.}

The $PIT$ histograms reveal problems with the Simple Average mixture, which match our discussion of the two regimes in Figures \ref{fig2} and \ref{fig:diff_heatmap_avg}, and which are  ameliorated  by the Best ${\leq} 4$-Average regularization. In particular, the Simple Average $PIT$ histograms show noticeable deviations from uniformity in both subsamples, and the shapes of the deviations are very different.

In the first subsample, the Simple Average $PIT$ histogram is highly skewed as shown in the upper-left panel of Figure \ref{aaabbb}, with far too little probability mass near 0 and far too much near 1, again indicating too many large inflation realizations relative to the Simple Average density forecasts. Regularization, however, shifts the densities upward as discussed earlier, producing an improved (if still imperfect) ${\leq} 4$-Average $PIT$ as seen in the bottom left panel of Figure \ref{aaabbb}. 

In the second subsample, the Simple Average $PIT$ histogram is more U-shaped as shown in the upper-right panel of Figure \ref{aaabbb}. In this regime the regularization spreads out the densities as discussed earlier, better accommodating the tail realizations and producing an improved Best ${\leq} 4$-Average $PIT$ as seen in the bottom right panel of Figure \ref{aaabbb}.

Finally, in parallel to our earlier examination of ECB/SPF inflation forecasts,  we examine real interest rate density forecasts. The real interest rate density is a simple sign change and location shift of the inflation density:
\begin{equation} \label{switch}
f(r_{t,t+1}) = i_{t,t+1} - f(\pi_{t,t+1}),
\end{equation}
where $r$ denotes the real interest rate, $i$ denotes the nominal interest rate, and $\pi$ denotes inflation.  Real interest rate densities are of course driven by the inflation densities via equation (\ref{switch}), but it is nevertheless interesting to make the translation into the real cost of borrowing. 

In Figure \ref{realint3D} we show the Simple Average and Best ${\leq}4$-Average  real interest rate density forecasts, and in Figure \ref{fig:diff_heatmap_avg realint} we show the differences between them, together with the realizations.\footnote{There is no need to show  regularized estimation results for real interest rates, because the log score is invariant to the switch from inflation to real interest rate density forecasts defined by equation (\ref{switch}). There is also no need to include Simplex panels in Figures \ref{realint3D} and \ref{fig:diff_heatmap_avg realint}, because the Simplex and Best ${\leq} 4$-Average inflation density regularizations, and hence real interest rate density regularizations,  are nearly identical. And finally, there is also no need to show real interest rate $PIT$ histograms, because they are exact mirror images of the inflation $PIT$ histograms in Figure \ref{aaabbb}, as revealed by equation (\ref{switch}).} One is immediately struck by the  high probability assigned to negative real rates through much of the sample.  $P(r_{t,t+1}){<}0$ is, for example, routinely greater than 1/2 since the end of the Great Recession, and the realized real rates often \textit{are} negative. 

Nevertheless our earlier inflation patterns and lessons remain firmly intact, because real interest rate density forecasts are driven by inflation density forecasts. There are two clear real interest rate ``regularization regimes," demarcated by the onset of the Great Recession. In the first, real interest rate densities are pushed downward, because, as discussed earlier,  regularization pushes inflation densities upward.  In the second, real interest rate densities are made more dispersed, because regularization makes inflation densities more dispersed.

\section{Concluding Remarks and Directions for Future Research}
\label{concl}

We have proposed methods for constructing regularized mixtures of density forecasts, exploring a variety of objectives and penalties, which we used in a  substantive exploration of Eurozone inflation and real interest rate survey density forecasts. All individual survey forecasters (even the ex post best  forecaster) are outperformed by our regularized mixtures. The log scores of the Simplex and Best-Average mixtures, for example, are approximately 7\% better than that of the ex post best individual forecaster, and  15\% better  than that of the ex post median forecaster.  Before the Great Recession, regularization shifts inflation density locations upward toward higher inflation, and hence real interest rate density locations downward, correcting for bias. From the Great Recession onward, the regularization tends to move probability mass from the centers to the tails of both inflation and real interest rate density forecasts, correcting for  overconfidence.

A variety of avenues for future research are possible. For example, one could use  the  probability integral transform  as a regularized  mixture estimation objective, minimizing a goodness-of-fit statistic (e.g., Kolmogorov-Smirnov) for testing the joint hypothesis of an $iid \, U(0,1)$  probability integral transform.  

Second, one could broaden our approach to allow for nonlinear mixtures as in recent work by \cite{McCalinn2020}, flexibly time-varying mixture weights as in \cite{JMV2010}, and mixture weights that  vary over regions of density support, as in \cite{Kape2015}.  

Finally, although we did not emphasize  regularization methods that require hyperparameter selection in our empirical work (Simplex+Ridge or Simplex+Entropy), they nevertheless represent interesting directions for future exploration.  An obvious issue is feasible real-time  hyperparameter selection.

\clearpage

\appendix
\appendixpage
\addappheadtotoc
\newcounter{saveeqn}
\setcounter{saveeqn}{\value{section}}
\renewcommand{\theequation}{\mbox{\Alph{saveeqn}.\arabic{equation}}} \setcounter{saveeqn}{1}
\setcounter{equation}{0}

\section{Derivation of the Simplex+Entropy Regularized Estimator}  \label{derivation}

The Simplex+Entropy estimator solves the  optimization problem: 
\begin{equation}
\label{log2a}
\hat{\omega} = \arg \min_{\omega} \left ( \underbrace{-\sum_{t=1}^{T} \log \left( \sum_{k=1}^{K} \omega_{k} f_{k,t}(y_{t})\right)}_{\text{log score}}+ \underbrace{(\alpha-1) \left(-\sum_{k=1}^{K} \log(\omega_{k})\right)}_{\text{entropy penalty} } \right )
\end{equation}
\begin{equation*}
 \text{s.t.}~ \omega_{k} \in (0,1), ~ \sum_{k=1}^{K}\omega_{k} = 1.
\end{equation*}
As we will show,  this it arises as the posterior mode in a  Bayesian  analysis with (1) log likelihood given by the log score, and (2) Dirichlet prior, which puts positive probability only on the unit simplex but also shrinks toward equal weights for a certain hyperparameter configuration. In particular, the $K$-dimensional Dirichlet prior is governed by $K$ hyperparameters, and when they equal, the prior mean is $1/K$. Hence the simplex+entropy regularization (\ref{log2}) with equal prior hyperparameters does the same thing as simplex+ridge (\ref{cup}): Impose simplex and shrink toward equal weights.

\subsection{Prior}

The  Dirichlet prior on ${\omega} =(\omega_{1}, \omega_{2}, ..., \omega_{K})$ with hyperparameter ${\boldsymbol \alpha} = (\alpha_{1}, \alpha_{2}, ..., \alpha_{K})$ is
\[
f_{D}(\omega; \boldsymbol \alpha) = \frac{1}{B(\boldsymbol  \alpha)} \prod_{k=1}^{K} \omega_{k}^{\alpha_{k}-1},
\]
where $B(\cdot)$ is the beta function, $\alpha_{k} >0 ~\forall k \in 1, ..., K$, and the support of $\omega$ is $\omega_{k} {\in} (0,1)$ with $\sum_{k=1}^{K} \omega_{k}=1$.

As is well known, the Dirichlet mean and variance are:

$$E(\omega_i) = \frac{\alpha_i}{\sum_{k=1}^K  \alpha_k}$$
and
$$
var(\omega_i) = \frac{
                \frac{\alpha_i}{\sum_{k=1}^K  \alpha_k}  \left  (1 - \frac{\alpha_i}{\sum_{k=1}^K  \alpha_k}  \right  )    }{1 + \sum_{k=1}^K  \alpha_k}.
        $$
Hence when $\alpha_{1} = \alpha_{2} = ... = \alpha_{K} = \alpha$, we have 
\[
E[\omega_{k}] = 1/K 
\]
and
\[
Var(\omega_{k}) = \frac{K-1}{\alpha K^{3} +\ K^{2}},
\]
for all $k = 1, ..., K$.  That is, the prior is centered on equal weights $1/K$, and  $var(\omega_{k}) {\rightarrow} 0$ as $\alpha {\rightarrow}  \infty$, so that $\alpha$ governs prior precision, with larger $\alpha$ producing heavier shrinkage toward $1/K$. 

\subsection{Posterior} 

The posterior distribution is 
\[
f_{D}(\omega | y; \boldsymbol \alpha) = \underbrace{
\prod_{t=1}^{T} \left( \sum_{k=1}^{K} \omega_{k} f_{k,t}(y_{t})\right)}_{\text{pseudo-likelihood}} \times \underbrace{\frac{1}{B(\boldsymbol\alpha)} \prod_{k=1}^{K} \omega_{k}^{\alpha-1}}_{\text{prior}},
\]
so the log posterior  is
\[
 \log f_{D}(\omega; \boldsymbol \alpha) = {\sum_{t=1}^{T} \log \left( \sum_{k=1}^{K} \omega_{k} f_{k,t}(y_{t})\right)}+ (\alpha-1) \sum_{k=1}^{K} \log(\omega_{k})-\log B(\boldsymbol\alpha).
\]
Because  $B(\boldsymbol\alpha)$ does not depend on $\boldsymbol \omega$, we can drop the last term, so the posterior mode is 
\begin{equation}
 \label{final}
\hat{\omega} = \arg \min_{\omega} \left ( \underbrace{-\sum_{t=1}^{T} \log \left( \sum_{k=1}^{K} \omega_{k} f_{k,t}(y_{t})\right)}_{\text{Log score}}+ \underbrace{(\alpha-1) \left(-\sum_{k=1}^{K} \log(\omega_{k})\right)}_{\text{penalty}}  \right )
\end{equation}
\[
 \text{s.t. } \omega_{k} \in (0,1), ~ \sum_{k=1}^{K}\omega_{k} = 1.
\]

\subsection{Understanding the Penalty Term}

 One way to understand the penalty term is to recall the solution to the empirical likelihood maximization problem of \cite{owen2001},
\[
\arg \min_{\omega} \left ( -\sum_{k=1}^{K} \log(\omega_{k}) \right )
\]
\[
\text{s.t. } \omega_{k} \in (0,1), ~ \sum_{k=1}^{K}\omega_{k} = 1,
\]
which is equal weights, $\omega_{k}{=}1/K,~ \forall k$. Hence we see that the penalty  part of (\ref{final}) is minimized at $\omega_{k}{=}1/K$, which yields a clear interpretation of the penalty term.
 Larger $\alpha$ means a tighter prior on $\omega$, with heavier shrinkage toward equal weights.    Several interesting limiting cases emerge.  First, for $\alpha {\rightarrow} \infty$, the penalty term dominates, and the optimal solution is equal weights.  Second, for $\alpha {\rightarrow} 1$, the penalty term vanishes, and the optimal solution matches that of the  optimal linear pool, with simplex constraint imposed. Third, there is a upper bound for $var(\omega_{k})$: as $\alpha {\rightarrow} 0$, $var(\omega_{k}) {\rightarrow} (K-1)/K^{2}$.

\subsection{Remarks}

\begin{enumerate}
        
\item The entropy regularization optimization problem is convex, because both the the log-score and the penalty  are convex.  A closed form may not exist for the regularized $\omega$, but  convexity makes numerical computation straightforward. 

\item Entropy regularization has a clear parallel to ridge regularization. As is well known, ridge regularization  emerges as the posterior mode in a Bayesian analysis with Gaussian prior, and as we have shown,  entropy regularization emerges as posterior mode in a Bayesian analysis with Dirichlet prior.   Both regularizations, moreover, are governed by a single parameter linked to prior precision. 

\item If the effects of the ridge and entropy penalties are very similar in certain respects (imposition of simplex and shrinkage toward $1/K$), their full Bayesian interpretations are nevertheless different.  In particular, the ridge (Gaussian) and entropy (Dirichlet) priors differ, even if their means are the same ($1/K$), and so the posteriors differ.  For $\alpha <1$ the Dirichlet prior distribution may not even have a single mode.

\end{enumerate}

\clearpage

\bibliographystyle{Diebold}
\addcontentsline{toc}{section}{References}
\bibliography{Bibliography}

\end{document}